\documentclass[prl,twocolumn,twoside,showpacs,superscriptaddress,floatfix]{revtex4-2}
\usepackage{graphicx,amssymb,amsmath,amsfonts,epsfig,color,enumerate,bm}
\usepackage{siunitx}
\usepackage[colorlinks=true, linkcolor=blue, citecolor=blue, urlcolor=blue]{hyperref}
\usepackage{algorithm,algpseudocode}
\usepackage[caption=false]{subfig}
\usepackage{float} 
\usepackage[percent]{overpic}
\usepackage[section]{placeins}
\usepackage{afterpage}
\makeatletter
\let\auto@bib@innerbib\@empty
\makeatother

\epsfclipon
\begin{document}
 \title{Cell State Transitions Beyond the Small-Noise Limit}
\author{Jianzhe Wei$^{\ddagger}$}
\affiliation{College of Enginneering, Huazhong Agricultural University, Wuhan 430070, China}
\affiliation{State Key Laboratory for Quantitative Synthetic Biology, Shenzhen Institute of Synthetic Biology, Shenzhen Institutes of Advanced Technology, Chinese Academy of Sciences, Shenzhen, 518055, China}
\email[J. W., J. Z., and P. C. contributed equally to this work.]{}
\author{Jingwen Zhu$^{\ddagger}$}
\author{Pan Chu$^{\ddagger}$}
\affiliation{State Key Laboratory for Quantitative Synthetic Biology, Shenzhen Institute of Synthetic Biology, Shenzhen Institutes of Advanced Technology, Chinese Academy of Sciences, Shenzhen, 518055, China}
\author{Liang Luo}
\email[Corresponding author.~]{luoliang@mail.hzau.edu.cn}
\affiliation{College of Enginneering, Huazhong Agricultural University, Wuhan 430070, China}
\author{Xiongfei Fu}
\email[Corresponding author.~]{xiongfei.fu@siat.ac.cn}
\affiliation{State Key Laboratory for Quantitative Synthetic Biology, Shenzhen Institute of Synthetic Biology, Shenzhen Institutes of Advanced Technology, Chinese Academy of Sciences, Shenzhen, 518055, China}

\begin{abstract}
State transitions are fundamental in biological systems but challenging to observe directly. Here, we present the first single-cell observation of state transitions in a synthetic bacterial genetic circuit. Using a mother machine, we tracked over 1007 cells for 27 hours. First-passage analysis and dynamical reconstruction reveal that transitions occur \textit{outside} the small-noise regime, challenging the applicability of classical Kramers' theory. The process lacks a single characteristic rate, questioning the paradigm of transitions between discrete cell states. We observe significant multiplicative noise that distorts the effective potential landscape yet \textit{increases} transition times. These findings necessitate theoretical frameworks for biological state transitions beyond the small-noise assumption.
\end{abstract}

\maketitle

 State transitions are fundamental to biological systems, enabling adaptation to environmental changes and driving cellular development and specialization. Experimental evidence from fluorescent imaging has traditionally revealed discrete switching in specific genetic circuits \cite{elowitz00,collins00,ozbudak04,wangxiao21,zhu23,chu25}. Such processes are frequently analyzed within the framework of Kramers’ transition state theory \cite{kramers40}. While the theory itself is broad, its application in the small-noise limit leads to the paradigm of discrete transitions characterized by a single, well-defined rate \cite{hanggi90,melnikov91}. While theoretical physics has long explored stochastic dynamics beyond this limit, including the effects of large or state-dependent noise, it remains an open question to what extent such non-classical regimes are relevant to functional biological circuits.

This possibility is supported by single-cell RNA sequencing (scRNA-seq) data, where cell state transitions often appear as a continuous "flow" through high-dimensional space \cite{manno18,wagner20,weinreb20}. From a physical standpoint, such a "flow" represents a case where potential barriers are low relative to the noise intensity, making the transition a continuous progression along a potential "valley" rather than a rare, impulsive jump. However, since static scRNA-seq snapshots cannot resolve real-time trajectories, the underlying dynamics of these "flows" remain unknown. Bridging this gap requires high-resolution temporal data at the single-cell level.

\begin{figure}[t]
  \centering
  \includegraphics[width=0.9\columnwidth]{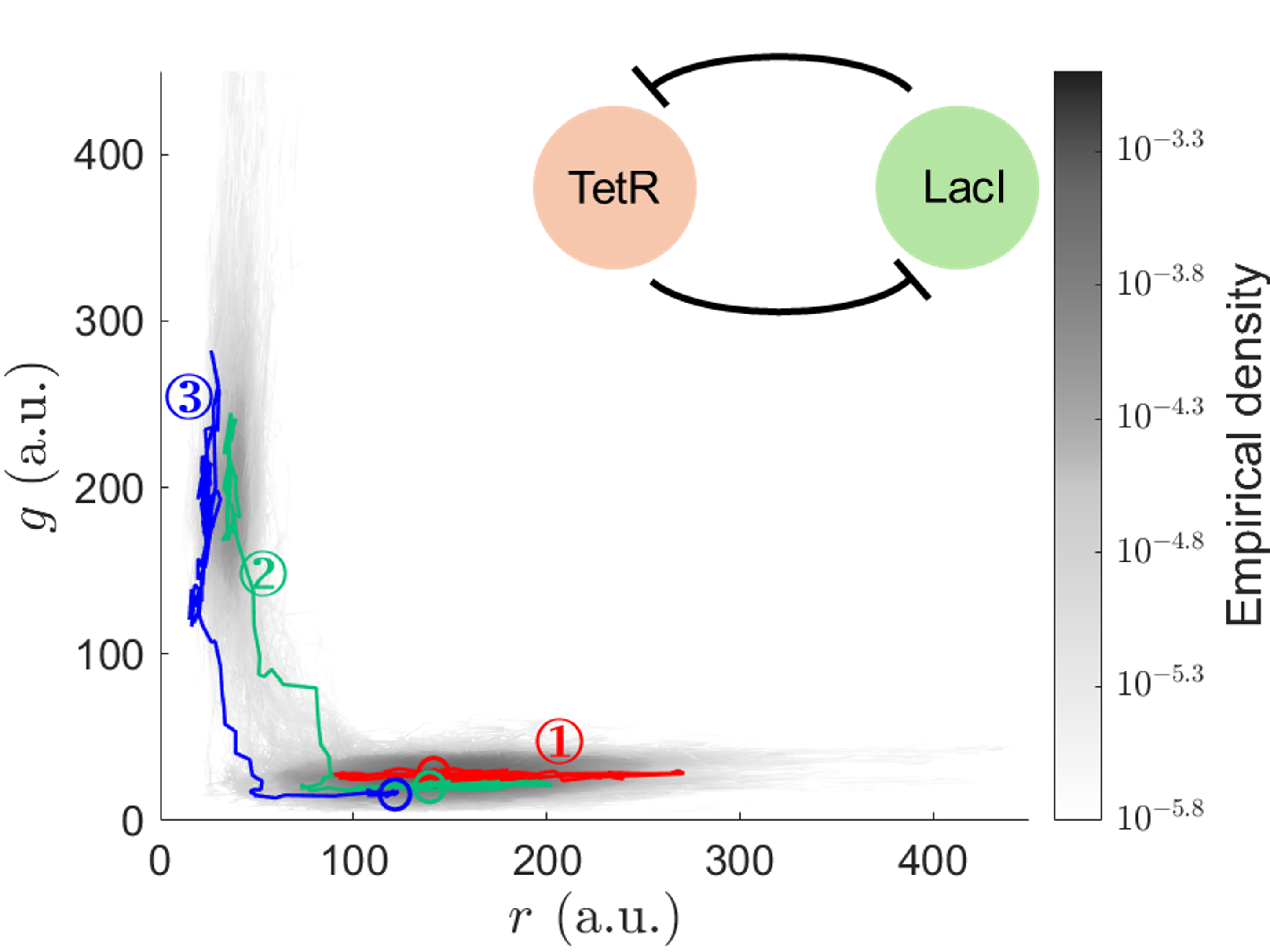}
  \caption{\label{fig1}  Temporal evolution of the synthetic toggle switch circuit (inset) is tracked at the single-cell level. All 1007 trajectories of RFP ($r$) and GFP ($g$) fluorescence intensities are plotted on the $r-g$ plane. Grayscale indicates the log-scale empirical density of $(r,g)$ readings over the entire observation period. Three representative trajectories are highlighted in color, and their initial positions are marked by open circles.
 }
\end{figure}

In this letter, we report direct observation of state transitions in a synthetic genetic toggle switch  using single-cell fluorescence imaging. Our results challenge the small-noise assumption, revealing strong noise that invalidates the single-rate transition paradigm. Through first-passage analysis, we further demonstrate that multiplicative noise prolongs transition times while effectively lowering the potential barrier height. This counterintuitive behavior stems from a localization mechanism analogous to quenched disordered systems, offering new perspectives on biological state transitions.

We study state transitions in an {\it Escherichia coli} toggle switch circuit (Fig.~\ref{fig1}) employing green (GFP) and red (RFP) fluorescent protein reporters \cite{elowitz00,zhu23,chu25}. The previous population-level studies demonstrate that this synthetic circuit—featuring two mutually repressive genes—exhibits bistability between green (G-state) and red (R-state) states under fast growth conditions, with asymmetric switching probabilities favoring R→G transitions \cite{zhu23}. 
In the current study, we tracked individual mother cells over extended periods using a mother machine, recording GFP and RFP relative intensities ($g(t)$ and $r(t)$) as time series.  
 This single-cell resolution enables first-passage analysis and dynamical reconstruction, both of which are inaccessible to population-level observations.

In our experiment, bacterial cells were first induced into the R-state using IPTG and loaded into a mother machine microfluidic chip. Following a $3$-hour acclimation in fresh MOPS‑buffered EZ rich defined medium (RDM) without IPTG, cells adapted to steady state growth phase ($\lambda = 1.6~\text{hr}^{-1}$) that favors the G-state. After the first three hours, 477 of the 1007 tracked mother cells retained the R-state. Their relaxation dynamics were observed over approximately 24 hours at $\Delta t = 0.1548$-hour resolution, yielding $N_{\text{t}} = 156$ frames per cell. 
(See Secs.~\ref{exp}--\ref{prep} in Supplementary Information (SI) for details.)

 Figure~\ref{fig1} displays all trajectories on the $r$-$g$ plane. Although incomplete transitions prevent ergodicity, the empirical density reveals a clear gap between the R-state and G-state. Non-parametric reconstruction\cite{vincent02,ohkubo11} of the two-dimensional dynamics yields drift streamlines that explicitly resolve the two attractors and their boundary (see Secs.~\ref{sec:2d_recon_hodge}--\ref{sec:2d_absorbing_boundary} in SI). Using this boundary as the transition threshold, approximately $62\%$ of the cells completed the first passage during the observation window.The mean and variance of $(r, g)$ remain stable for R-state cells but continuously evolve for G-state cells due to ongoing post-transition relaxation. We therefore focus exclusively on the first-passage process from the R-state to the boundary, deferring the analysis of subsequent relaxation dynamics.

The first-passage time (FPT) $\tau$ is central to transition state theory, defined as the time for a system to evolve from an initial position $\{r_0, g_0\}$ within an attractor to a specified boundary. Under timescale separation assumptions, the transition rate $k$ equals the inverse mean FPT ($k = \langle \tau \rangle^{-1}$).
The single-cell experimental data enable the direct measurement of the cell-specific first-passage times $\tau(r_0, g_0)$ and their dependence on initial conditions.  Surprisingly, we observe fold differences in $\tau$ across different initial positions $\{r_0, g_0\}$. This substantial variation persists despite all cells starting from the same  R-state and experiencing identical environmental conditions. 
This finding has profound implications: The process cannot be adequately understood as a simple transition between two discrete states, as implied by the phrase “from R-state to G-state”. At least, it is not a single-step process characterized by a uniform transition rate. Instead, the data indicate that the local relaxation timescale within the attractor is not well separated from the barrier-crossing timescale,  directly challenging the small-noise assumption widely adopted in classical Kramers' transition state theory.

To quantitatively validate this hypothesis, we proceed to reconstruct the effective dynamics from the experimental data. The main challenge lies in the limited statistical sampling. While the non-parametric estimation captures the structure of the attractors, the current ensemble of trajectories is insufficient to resolve the full two-dimensional stochastic dynamics in the $r-g$ space with adequate resolution. Fortunately, we observe that in the R-state, $g$ remains steadily suppressed within a narrow range and exhibits minimal fluctuations. This experimental observation justifies approximating the dynamics through an effective one-dimensional reduction along the $r$-dimension, effectively integrating out the fluctuations in the $g$-dimension. This reduced dynamics can be expressed in the form of a Fokker-Planck equation:
\begin{equation}
\label{fpe}
\frac{\partial P(r,t)}{\partial t} = -\frac{\partial}{\partial r}\left[ f(r) P(r,t) \right] + \frac{\partial^2}{\partial r^2} \left[ D(r) P(r,t) \right],
\end{equation}
where $f(r) = -dU(r)/dr$ represents the drift force derived from a potential landscape $U(r)$. 
Following well-established methods~\cite{ching99,tang03,friedrich11,ao13}, we estimate $f(r)$ and $D(r)$ from the displacement $\Delta r = r(t+\Delta t) - r(t)$ using:
\begin{align}
\label{recf}
f(r) &= \frac{1}{\Delta t} \left\langle r(t+\Delta t) - r(t) \right\rangle \vert_{r(t)=r}, \\
\label{recd}
D(r) &= \frac{1}{2\Delta t} \left\langle \left| r(t+\Delta t) - r(t) - f(r) \Delta t \right|^2 \right\rangle \vert_{r(t)=r},  
\end{align}
where $\langle \cdot \rangle \vert_{r(t)=r}$ denotes the ensemble average over all increments originating from position $r$.

\begin{figure}
   \centering
   \includegraphics[width=8cm]{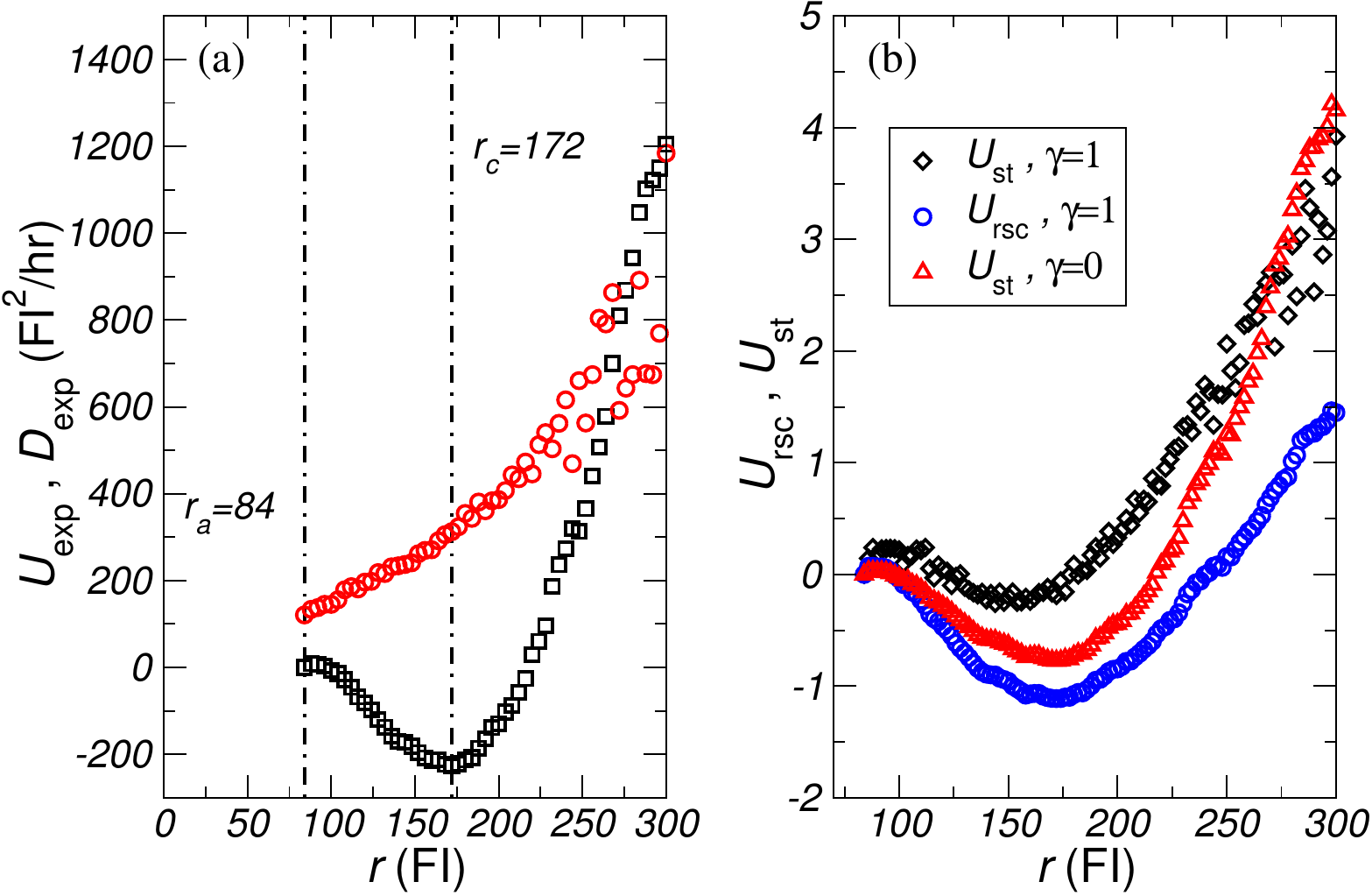} 
   \caption{(a) Reconstructed landscape $U$ (black squares) and the noise strength $D$ (red circles) from the experiment data. Both quantities share the same dimensions with the unit $\text{FI}^2/\text{hr}$. The unit of $r$ is FI, the fluorescent intensity. $U$ and $D$ are estimated for the $R$-state regime with the boundary $r_a=84$. The center of the trap locates at $r_c=172$. (b) Rescaled landscape $U_{\text{rsc}}$, defined in Eq. (\ref{rsc}) (blue circles), and the modified one $U_{\text{st}}=\ln D+U_{\text{rsc}}$ (black squares). The red triangles show $U_{\text{st}}$ from a case of homogeneous noise as comparison. (See Eq.(\ref{dgamma}) and Fig. \ref{fig3}(b) for the definition of $\gamma$. )  } 
   \label{fig2}
\end{figure}
  
Focusing on first passage to the barrier near $r_a = 84$, we estimate $f(r)$ and $D(r)$ for $r > r_a$ (see Sec.~\ref{rec} in SI). As shown in Fig.~\ref{fig2}(a), the reconstructed potential landscape $U(r) = -\int_{r_a}^{r} f(r')  dr'$ exhibits a characteristic barrier-trap structure. The diffusion coefficient $D(r)$ displays significant position dependence, confirming multiplicative noise. Crucially, the scale of variations in $U(r)$ is comparable to the magnitude of $D(r)$. 
Recognizing the intrinsic coupling between barrier height and noise strength in multiplicative noise systems, we define a rescaled landscape:
\begin{equation}
\label{rsc}
U_{\text{rsc}}(r) = -\int_{r_a}^{r} dr'  \frac{f(r')}{D(r')}. 
\end{equation} 
Additionally, we consider a modified landscape derived from the stationary distribution $P_{\text{st}}$ \cite{rosas16,moreno20,commer22}: $U_{\text{st}}(r) = -\ln P_{\text{st}}(r)$. Using the relation $P_{\text{st}}(r) = N_0 e^{-U_{\text{rsc}}(r)} / D(r)$ \cite{risken}, we obtain:
\begin{equation}
U_{\text{st}}(r) = U_{\text{rsc}}(r) + \ln D(r).
\end{equation}
Figure~\ref{fig2}(b) compares $U_{\text{rsc}}$ and $U_{\text{st}}$. Both landscapes exhibit shallow barrier-trap structures that contrast sharply with the deep-well assumptions of small-noise-limit theories. This implies comparable timescales for intra-trap relaxation and barrier crossing. Consequently, when initial positions $r_0 = r(t=0)$ are not fully relaxed within the trap, FPT exhibit strong $r_0$-dependence.

\begin{figure}
   \centering
   \includegraphics[width=8cm]{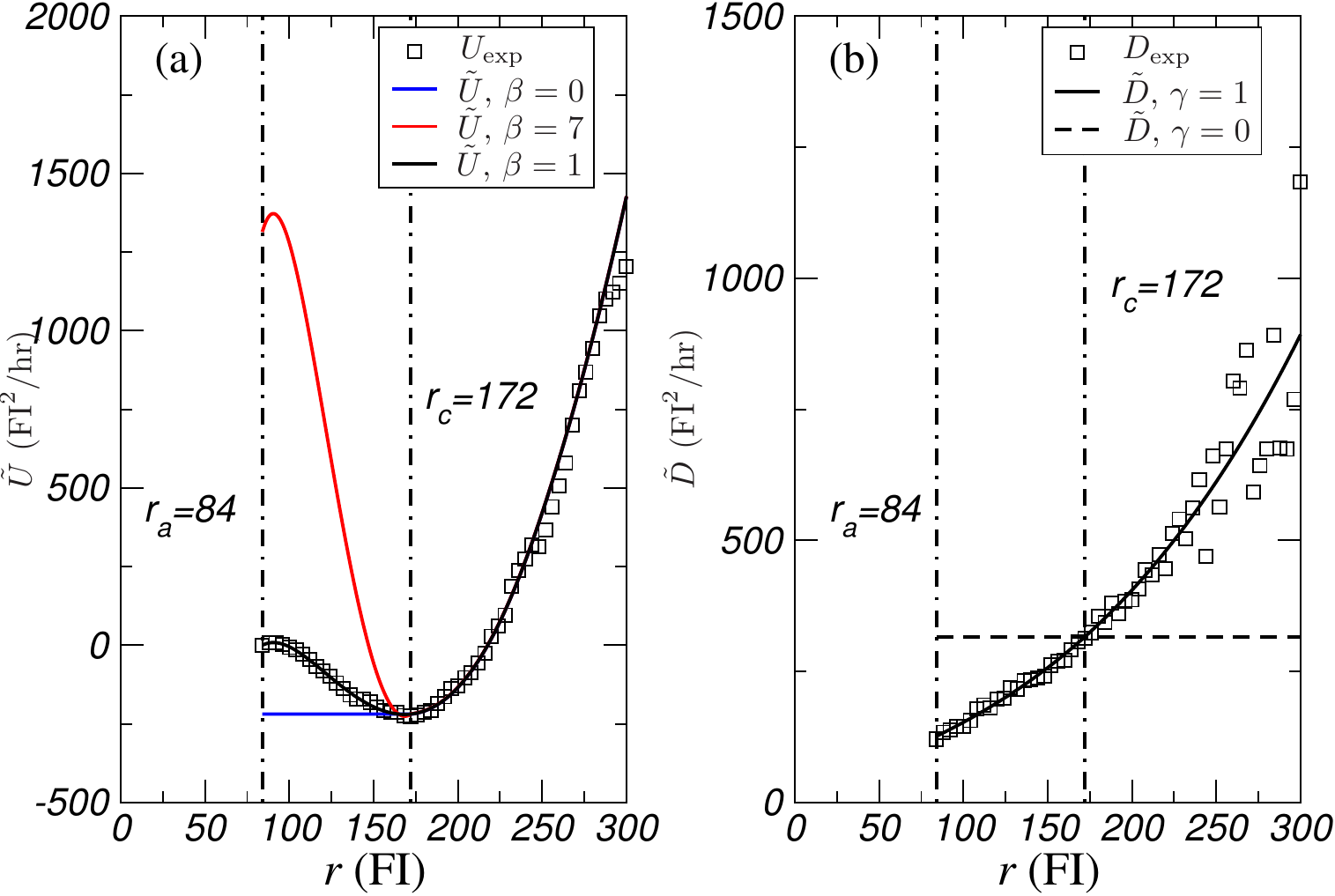} 
   \caption{Modified landscape $\tilde{U}$ and noise strength $\tilde{D}$ used in simulations. (a) The landscape $\tilde{U}(r;\beta)=-\int dr' \tilde{f}(r';\beta)$, calculated using $\tilde{f}$ from Eq.~(\ref{fbeta}), is shown as solid lines. Squares indicate the original landscape reconstructed from experimental data. (b) The noise strength $\tilde{D}(r;\gamma)$, defined by Eq.~(\ref{dgamma}), is shown as lines. Squares represent the original noise strength from experiment. } 
   \label{fig3}
\end{figure}

 First-passage times are determined by simulating the first-passage process of the reconstructed dynamics, as the experimental data lack sufficient sampling for high-resolution statistical analysis. For the simulation, the noisy experimental $U_{\text{exp}}$ and $D_{\text{exp}}$ are fitted by smooth functions $U_{\text{fit}}$ and $D_{\text{fit}}$. To systematically investigate the separate contributions of the potential landscape and multiplicative noise, we introduce two control parameters $\beta$ and $\gamma$ that modulate the dynamics. The parameter $\beta$ reshapes the landscape continuously via the modified drift force as
\begin{equation}
\label{fbeta}
\tilde{f}(r;\beta)=f_{\text{fit}}(r)\left[\beta H(r_c-r)+H(r-r_c)\right],
\end{equation}
where $H(x)$ is the Heaviside step function. This transformation rescales the landscape between the trap center $r_c$ and the absorbing boundary while preserving it for $r > r_c$, as shown in Fig.~\ref{fig3}(a). When $\beta=0$, the landscape has no barrier. For $\beta>1$, the barrier height increases to $\Delta \tilde{U}=\beta\Delta U$, corresponding to the small-noise limit. The original fitted landscape is recovered when $\beta=1$, giving $\tilde{U}=U_{\text{fit}}$.
The parameter $\gamma$ tunes the $r$-dependence of the noise strength by
\begin{equation}
\label{dgamma}
\tilde{D}(r;\gamma)=D(r_c)+\gamma\left[D_{\text{fit}}(r)-D(r_c)\right].
\end{equation}
For $\gamma=0$, the noise strength is homogeneous for all $r$, with $D=D(r_c)$. As $\gamma$ increases, the $r$-dependence is gradually reintroduced. The original fitted profile is recovered when $\gamma=1$, giving $\tilde{D}=D_{\text{fit}}$, as shown in Fig.\ref{fig3}(b).

\begin{figure}
   \centering
   \includegraphics[width=8cm]{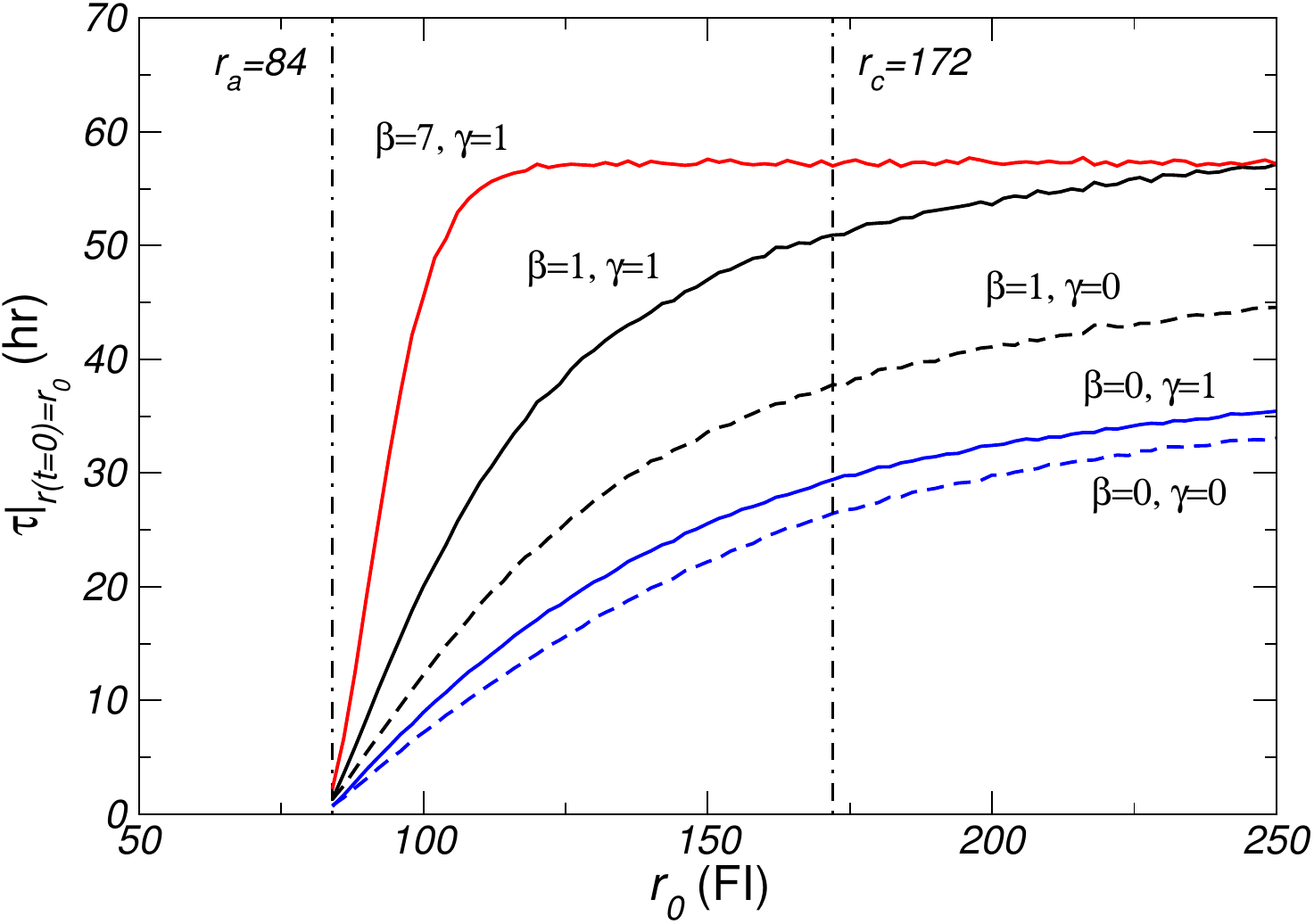} 
   \caption{Mean first-passage time (MFPT) $\tau$ to the boundary at $r_a=84$ for the initial value $r(t=0)=r_0$, simulated by Langevin dynamics using the reconstructed $f(r)$ and $D(r)$, illustrating the contributions from the landscape and the multiplicative noise. Line colors indicate different barrier scales (see Eq.~(\ref{fbeta})), consistent with Fig.~\ref{fig3}(a). Solid lines correspond to the observed case of multiplicative noise ($\gamma=1$), while dashed lines represent the case of homogeneous noise intensity ($\gamma=0$, see Eq.~(\ref{dgamma})), consistent with Fig.~\ref{fig3}(b). The MFPT of the small-noise case with $\beta=7$ (red solid line) is significantly larger ($\tau(r_0=250)\simeq 1.2 \times 10^4$ hr) and has been rescaled here for better visualization.}
   \label{fig4}
\end{figure}

The mean first-passage time (MFPT) $\tau$ from initial positions $r_0$ is shown in Fig.~\ref{fig4} for various $\beta$ and $\gamma$ values. In the small-noise limit ($\beta=7$), $\tau$ increases sharply, soon plateauing at $\tau_k\simeq 1.2\times 10^4 \text{hr}$. This behavior allows classical transition rate theory to characterize the process with a single transition rate $r_k = 1/\tau_k$, independent of the specific initial value $r_0$. In contrast, the original dynamics ($\beta=1$, $\gamma=1$) exhibit a gradual increase in $\tau$ across a wide range of $r_0$ values—from the absorbing boundary $r_a$ to the trap center $r_c$ where most cells reside. This clear deviation from small-noise behavior demonstrates the inapplicability of the single-rate approximation for this system.

Multiplicative noise is ubiquitous in biological systems with complex regulatory mechanisms. As shown in Fig.~\ref{fig2}(a), the reconstructed dynamics exhibit strong multiplicative noise, where the noise strength $D(r)$ increases with $r$. To evaluate the kinetic effect of this state-dependence, we use Eq.~(\ref{dgamma}) to simulate and compare the dynamics under the original multiplicative noise ($\gamma=1$) and additive noise ($\gamma=0$) (see Fig.~\ref{fig3} for the original and modified noise strength). As shown in Fig.~\ref{fig4}, the presence of multiplicative noise prolongs the mean first-passage time.

As established theoretically, multiplicative noise alters the stationary distribution $P_{\text{st}}$~\cite{commer22}, which is reflected in the modified effective landscape $U_{\text{st}}=-\ln P_{\text{st}}$. However, Fig.~\ref{fig2}(b) shows that multiplicative noise actually lowers the barrier in $U_{\text{st}}$. This apparent reduction in barrier height contradicts the observed increase in transition times. This paradox arises from the difference between long-time steady-state statistics and transient escape dynamics. Multiplicative noise biases the stochastic trajectory to spend more time in regions with lower noise intensity. In the long-time limit, the cellular state repeatedly samples the attractor, and the stationary distribution, derived from occupation time, shifts toward regions of weaker noise. In contrast, the state transition is a first-passage process. During trap escape from the trap center $r_c$ to the barrier peak, the decreasing noise strength extends the transition time. This indicates that theoretical frameworks built upon the long-time limit may not be applicable to dynamic processes such as cell-state transitions.

 In this letter, we have reported the direct observation of state transitions in a synthetic genetic toggle switch at the single-cell level. Over the past decades, state transition theory within the small-noise limit has seen significant development, from refined calculations of transition rates to the reconstruction of Waddington landscapes from ensemble data~\cite{hanggi90,melnikov91,ao04,ao17,wang19}. However, our results show that genetic circuits can operate in a regime where noise is comparable to barrier heights, rendering traditional small-noise approximations and single-rate descriptions inadequate. These findings bring renewed focus to the question of whether state transitions are primarily stochastic, noise-driven jumps over barriers, or if the intervention of signals actively reshapes the landscape to eliminate such barriers. The latter scenario aligns more closely with biological intuition, and our synthetic system provides a concrete demonstration of this mechanism.

The strong dependence of transition times on initial conditions suggests that the conceptual definition of a cellular “state” requires reconsideration ~\cite{cellsyst17,mulas21}. Even when considering complex noise structures, such as colored or bounded fluctuations that can dynamically perturb the potential landscape~\cite{shahrezaei08,hsu16,donofrio18}, the existing framework remains centered on the paradigm of discrete switching between attractors. When the separation of timescales between intra-state fluctuations and inter-state transitions disappears, this paradigm may be replaced by a continuous “flow” through state space~\cite{ao04-2,wangjin07,stumpf17,weinreb18,wagner20,xing22}. This perspective is consistent with current interpretations of single-cell RNA sequencing data, where cell differentiation is often modeled as a continuous progression along a potential “valley” rather than an impulsive jump ~\cite{manno18,wagner20,weinreb20}. Our work highlights the need for theoretical developments beyond the small-noise limit. A comprehensive framework is called for to bridge the continuous flow inferred from static transcriptomics and the real-time stochastic dynamics revealed by continuous tracking.

\begin{acknowledgements}
This work is partially supported by the National Key Research and Development Program of China (2024YFA0919600), Strategic Priority Research Program of Chinese Academy of Sciences (XDB0480000), and NSFC (32301225, T2525031). 
\end{acknowledgements}

\clearpage
\begin{onecolumngrid}
\setcounter{section}{0}
\setcounter{figure}{0}
\setcounter{table}{0}
\setcounter{equation}{0}
\setcounter{page}{1} 

\renewcommand{\thesection}{S\arabic{section}}
\renewcommand{\thefigure}{S\arabic{figure}}
\renewcommand{\thetable}{S\Roman{table}}
\renewcommand{\theequation}{S\arabic{equation}}
\renewcommand{\thepage}{S\arabic{page}}
\renewcommand{\citenumfont}[1]{S#1}
\renewcommand{\bibnumfmt}[1]{[S#1]}
\nocite{wang2010robust,neidhardt1974culture,pachitariu2022cellpose,rang2018minicells,sutterlin2016disruption,french2017bacteria,benichou2010geometry,zhu2023unbalanced,cameron2014tunable,ohkubo2011nonparametric,vincent2002manifold,elvira2022rethinking}

\makeatletter
\renewcommand{\c@secnumdepth}{0}
\makeatother
\begin{center}
\Large\textbf{Supplementary Information for \\``Cell State Transitions Beyond the Small-Noise Limit''}
\end{center}
\vspace{1cm}

In this Supplemental Material, we provide technical details supporting the main text. Section~\ref{exp} describes the experimental setup and protocol. Section~\ref{prep} details the data preprocessing steps. Section~\ref{gr} analyzes the cellular growth rate to confirm physiological stability. Section~\ref{sec:2d_recon_hodge} presents the two-dimensional nonparametric reconstruction and the Helmholtz--Hodge decomposition of the reconstructed drift field. Section~\ref{sec:2d_absorbing_boundary} describes the selection of the absorbing boundary from the reconstructed two-dimensional dynamics. Section~\ref{fp} presents the first-passage analysis of single-cell data. Section~\ref{rec} presents the effective one-dimensional reconstruction and associated consistency checks. Section~\ref{sim} shows the details of the first-passage simulations and the resulting first passage time distributions. Additional supplementary figures are included at the end to further elucidate the experimental observations.

\vspace{2em}

\section{Experiment setup and Protocol}
\label{exp}

\subsection{Microfluidic Device (Mother Machine)}

A custom-designed PDMS microfluidic device ("mother machine") was used to monitor long-term bacterial growth, based on the design reported by Wang et al.~\cite{wang2010robust}. 
The device comprises an array of narrow side channels orthogonally connected to a main trench that continuously delivers fresh medium. Each growth channel was 20 or 25 $\mu$m in length, 1.0–1.5 $\mu$m in width, and approximately 1.0–1.2 $\mu$m in height. The main trench measured 25 $\mu$m in depth and 100 $\mu$m in width.

To fabricate the mother machine chips, polydimethylsiloxane (PDMS; Dow Corning, SYLGARD 184 Silicone Elastomer Kit) was prepared by thoroughly mixing the base and curing agent at a 10:1 (w/w) ratio. The mixture was degassed under vacuum (-0.8 kg/cm²) for 10 minutes, poured onto a patterned silicon wafer, and further degassed to remove surface bubbles. The PDMS was then cured at $\qty{80}{\degreeCelsius}$ for at least 30 minutes. After curing, the PDMS layer was demolded, cut into individual chips, and inlet and outlet holes (0.7 mm diameter) were punched. Cleaned glass coverslips (thickness 0.13–0.16 mm) were bonded to the feature side of the PDMS chips using oxygen plasma treatment (Harrick Plasma, PDC-32G) for 2 minutes, followed by incubation at $\qty{80}{\degreeCelsius}$ for at least 10 minutes to reinforce bonding.

\subsection{Bacterial Strain and Plasmid}

All experiments were conducted using \textit{Escherichia coli} strain derived from the K-12 NCM3722 background. The engineered strain NH3 was constructed by deleting the \textit{fliC} gene, encoding the flagellar structural protein, and the \textit{lac} operator. The wild-type NCM3722 strain was generously provided by Dr. Chenli Liu. Mutual repression gene circuits were introduced via the plasmid pECJ3 (Addgene plasmid \#75465, a gift from Dr. James Collins)~\cite{cameron2014tunable}, carried on a ColE1 origin plasmid backbone. The circuit consists of two mutually repressive transcription factors: LacI, expressed from the PLtetO-1 promoter, and TetR, expressed from the Ptrc2 promoter. Two distinct fluorescent reporters, GFPmut2 (GFP) and mCherry (RFP), respectively indicate the two opposing states. Under steady-state growth in nutrient-rich media (e.g., RDM), cells can be induced into either a green state (high LacI/GFP expression) or a red state (high TetR/RFP expression) by the appropriate chemical inducers. Once established, these states are stably maintained even after the removal of the inducers.

\subsection{Growth Medium and Cell Culture}

Cells were cultured in MOPS-buffered EZ rich defined medium (RDM)~\cite{neidhardt1974culture}, supplemented with 0.4\% (w/v) glucose and 10 $\mu$g/mL kanamycin to maintain plasmid selection. The nitrogen source was 9.5 mM $\text{NH}_4\text{Cl}$. Chemical inducers included IPTG (isopropyl $\beta$-D-1-thiogalactopyranoside; Sigma-Aldrich, I6758) at 0.2 mM and chlorotetracycline hydrochloride (cTc; Aladdin, C103023) at 10 ng/mL, used as needed to induce transitions to the red and green states, respectively.

Strains were initially streaked on LB agar plates from glycerol stocks stored at $\qty{-80}{\degreeCelsius}$ and incubated at $\qty{37}{\degreeCelsius}$ for 10–12 hours. Subsequently, 3–5 single colonies were selected and inoculated into 14 mL tubes containing 3 mL RDM medium. Cultures were grown overnight in a shaker (220 rpm, $\qty{37}{\degreeCelsius}$; Shanghai Zhichu Instrument) to generate seed cultures.
For pre-culture, seed cultures were diluted into fresh RDM medium supplemented with 0.2 mM IPTG to maintain cells in the red state, with an initial $\text{OD}_{600}$ of approximately 0.01. Pluronic F-108 (Sigma-Aldrich, 542342-250G) was added at a final concentration of 0.85 g/L to minimize biofilm formation. Successive dilutions were performed when $\text{OD}_{600}$ reached 0.2, repeating for several rounds to ensure balanced growth for at least 10 generations and establish steady-state conditions. Pre-cultures were maintained in a water-bath shaker (150 rpm, $\qty{37}{\degreeCelsius}$; Shanghai Zhichu Instrument) using 29 mm × 115 mm test tubes with no more than 10 mL of medium per tube. Cells from the final round of pre-culture were grown to an $\text{OD}_{600}$ of ~0.5 before loading into microfluidic devices.

Cultures were centrifuged and concentrated 100-400-fold, then loaded into mother machine chips and centrifuged at 2500$\times$g for 5 minutes to trap cells into side-channels. Fresh RDM medium without IPTG was perfused at a high flow rate for 10 minutes to clear blockages, after which the flow rate was reduced and maintained at 10 $\mu$L/min using a pressure controller (FluidicLab PC1) equipped with a 0.22 $\mu$m filter. Cells were allowed to equilibrate under continuous perfusion for 2–3 hours before imaging. The chip was mounted on a microscope stage equipped with a custom temperature control system set to $\qty{37}{\degreeCelsius}$ and humidity control maintained at approximately 60\%.

\begin{figure}
\includegraphics[width=0.8\textwidth]{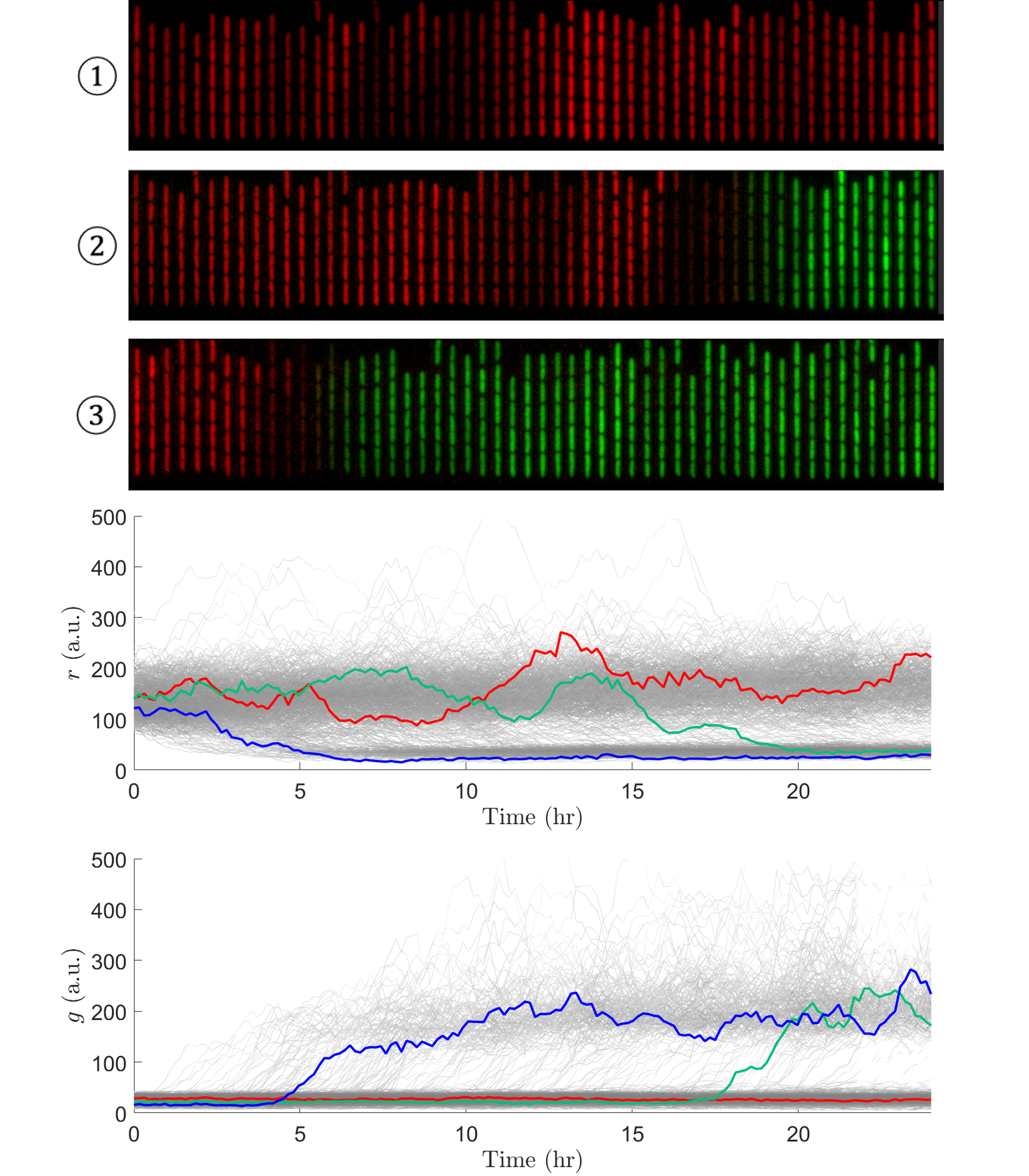}
\caption{Long-term single-cell tracking using the ``mother machine'' microfluidic device. The top three panels show $xy$--$t$ montages of raw fluorescence images for three representative growth channels, labeled 1 (red), 2 (green), and 3 (blue). The lower two panels show the corresponding time series of RFP intensity $r$ and GFP intensity $g$, respectively, together with those of all tracked mother cells growing in rich defined medium (RDM). These three representative cell lineages correspond exactly, in both numbering and color coding, to the three representative trajectories shown in Fig.~\ref{fig1} of the main text. Images were sampled every 3 frames, corresponding to about $9.3$ minutes per frame.}
\end{figure}

\subsection{Microscopy and Time-lapse Imaging}

Microscopic imaging was performed using a Nikon Ti-E inverted microscope equipped with a SpectraX LED light source (Lumencor) for epifluorescence illumination. A 100$\times$ oil immersion objective (Nikon Plan Apo $\lambda$, NA 1.5) was used for high-resolution single-cell tracking in the mother machine. Images were acquired with an ORCA-Flash4.0 sCMOS camera (Hamamatsu). Fluorescence signals from GFP and RFP were captured using a dual-band filter set (Chroma 59022).
For single-cell time-lapse experiments, phase-contrast images were acquired every 3 minutes, while GFP and RFP fluorescence channels were captured every 9 minutes over a total imaging period of 24–30 hours.

\subsection{Cell Segmentation and Single-Cell Tracking}

Custom image analysis pipelines incorporating deep learning algorithm Cellpose~\cite{pachitariu2022cellpose} were developed to process time-lapse data acquired from the mother machine. The workflow consisted of four primary steps:

\begin{enumerate}
\item Image Registration and Channel Detection: Time-lapse images from each field-of-view (FOV) were first registered to correct for XY drift caused by stage movement. A pre-trained model was used to identify and segment side channels within each FOV.

\item Cell Segmentation: Individual cells were segmented using the re-trained segmentation model, which was trained by our own mother machine data to recognize bacterial morphology. Edge refinement was performed using Otsu’s thresholding algorithm to enhance cell boundary detection.

\item Cell Geometry Extraction: Cell midlines were calculated via interpolation to provide initial estimates of cell geometry. From these, cell parameters—including mask, length, width, and area—were extracted using a channel-aligned coordinate system.

\item Fluorescence Quantification: Fluorescent protein expression levels were quantified by applying the segmented masks to fluorescence images. Background fluorescence was estimated using the median pixel intensity of each channel. The cellular fluorescence signal was computed by subtracting the background from the median intensity of pixels within each cell mask.
\end{enumerate}

\section{Data Preprocessing}
\label{prep}

This section describes the preprocessing steps for time-series data from the mother machine setup.

Occasionally, cells in the mother machine enter abnormal physiological states where the genetic circuit dynamics differ significantly from normal cells. These abnormal cells are identified through morphology and fluorescent intensity measurements. After excluding abnormal cells, the dataset contains $(r, g)$ trajectories for 1,007 cells, each spanning 27 hours (176 frames). Here, $r$ and $g$ represent the relative intensities (RI) of red fluorescent protein (RFP) and green fluorescent protein (GFP), respectively. This dataset serves as the basis for subsequent analysis and theoretical modeling. The following subsections detail the preprocessing steps.

To avoid confusion, we note that data analysis suggests the dynamics of the first 3 hours are significantly different from those of the later hours (see Fig.~\ref{fig:S6-2}(a)). This indicates that the cells are not in a steady state during the early experimental period. The first-passage analysis and the reconstructed dynamics in the main text involve only the later 24-hour data, which includes 477 cells. More details are discussed in Sec.~\ref{rec}. 

\subsection{Morphology-Based Filtering}

Physiological abnormalities are reflected in abnormal morphological features. These abnormalities are identified based on cell area and diameter measurements.

First, cells exhibiting extremely small areas are filtered. As shown in Fig.~\ref{fig:S2-1}(a), a small portion of the "cells" identified by the automatic segmentation algorithm have very small areas. These "minicells" are actually outer membrane blebs, which are widely recognized as resulting from accumulated damage, potentially linked to oxidative stress~\cite{rang2018minicells,sutterlin2016disruption}. To exclude minicells, any cell with a measured area below $0.4\:\mu\mathrm{m}^2$ is filtered. The threshold is marked as red dashed line in Fig.~\ref{fig:S2-1}(a). 

Next, cells with abnormally small diameters are filtered. Some cells identified by the automatic segmentation algorithm have reasonable areas but excessive length, resulting in diameters significantly smaller than those of normal cells. These shape abnormalities likely arise from altered metabolic activity or stress responses~\cite{french2017bacteria}. The cells with width less than $0.3\mu\text{m}$ are excluded from the analysis. The threshold is shown as red dashed line in Fig.~\ref{fig:S2-1}(b).

\begin{figure}[htbp]
\begin{overpic}[width=0.48\textwidth]{"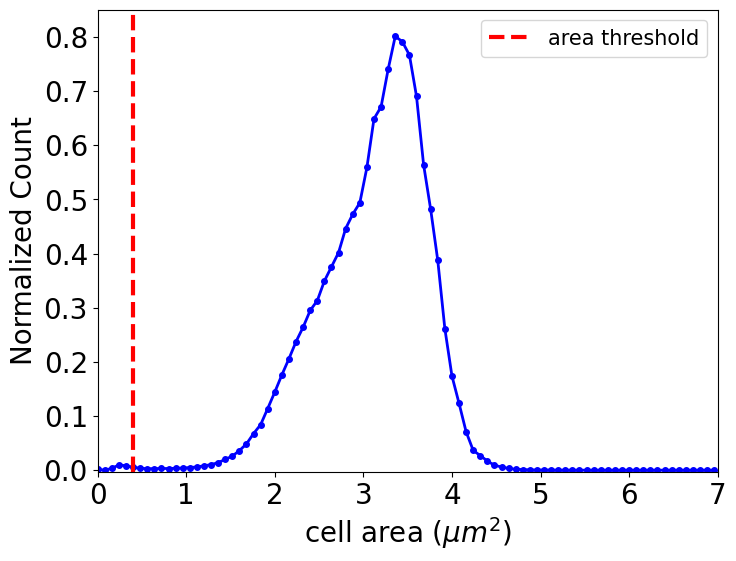"}
	 \put(-4,70){\Large\bfseries (a)}
\end{overpic}
\begin{overpic}[width=0.46\textwidth]{"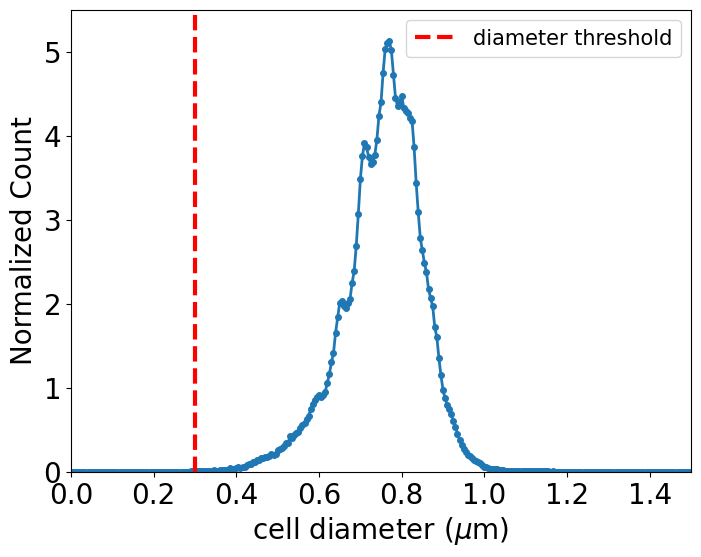"}
	\put(-3,73){\Large\bfseries (b)}   
\end{overpic}
	\caption{
    \label{fig:S2-1}
  Cell morphology statistics. Distribution of cell area (a) and inferred diameter (b) for all cells across all time frames. }
\end{figure}

\subsection{Fluorescence-Based Filtering}
\label{subsec:fluorescence_filtering}

Physiological abnormalities can significantly alter the dynamics of the genetic circuit, which manifests in abnormal fluorescence patterns. These abnormalities are identified and filtered based on fluorescence trajectory behavior.

Occasionally, cells enter a growth-stalled state where expression of most genes ceases, including those in the synthetic circuit. In these cases, both $r$ and $g$ readings remain low for long time.  In the $(r, g)$ plane shown in Fig.~\ref{fig:S2-2}, the fluorescence trajectory can stay in the lower-left corner for hours. These cells are excluded from the transition process statistics to focus on normal physiological state. 

In rare cases, fluorescent readings for both $r$ and $g$ simultaneously rise to very high values. This regulatory failure of plasmid-carried synthetic genes (unlike well controlled native genes) reflects the absence of robust regulation mechanisms. Such failures also cause occasional large jumps in fluorescent readings. Since these abnormal cases follow completely different circuit dynamics, their fluorescence trajectories are excluded from analysis. 

According to the above reasons, we filter the fluorescence trajectories following the below criterion. 
Let $\{r_i(t),g_i(t)\}$ denote the red and green fluorescence intensity of cell $i$ at frame $t$, representing its fluorescence trajectory over the set of time frames $t \in T$, where $T = \{1,2,\dots,176\}$ . Each frame in the time series corresponds to 0.1548 hours. Based on these trajectories, we removed cells satisfying any of the following conditions:
\begin{eqnarray}
S_1 &= &\{\,i \mid \exists\,t_0\in T \text{, such that } r_i(t)<45,\,g_i(t)<45\ ,\forall\,t\in[t_0,t_0+24]\,\},\label{eq:S1}\\
S_2 &= &\{\,i \mid \exists\,t\in T \text{, such that } r_i(t)>100,\,g_i(t)>70\,\},\label{eq:S2}\\
S_3 &= &\{\,i \mid \exists\,t\in T \text{, such that } \lvert \Delta r_i(t)\rvert > 50,\ \Delta r_i(t)=r_i(t)-r_i(t-1)\,\}.\label{eq:S3}
\end{eqnarray}
The red lines in Fig.~\ref{fig:S2-2} indicate the regions. Cells belonging to \(S_1 \cup S_2 \cup S_3\) were filtered. 

\begin{figure}[htbp]
    \centering
    \includegraphics[width=0.6\textwidth]{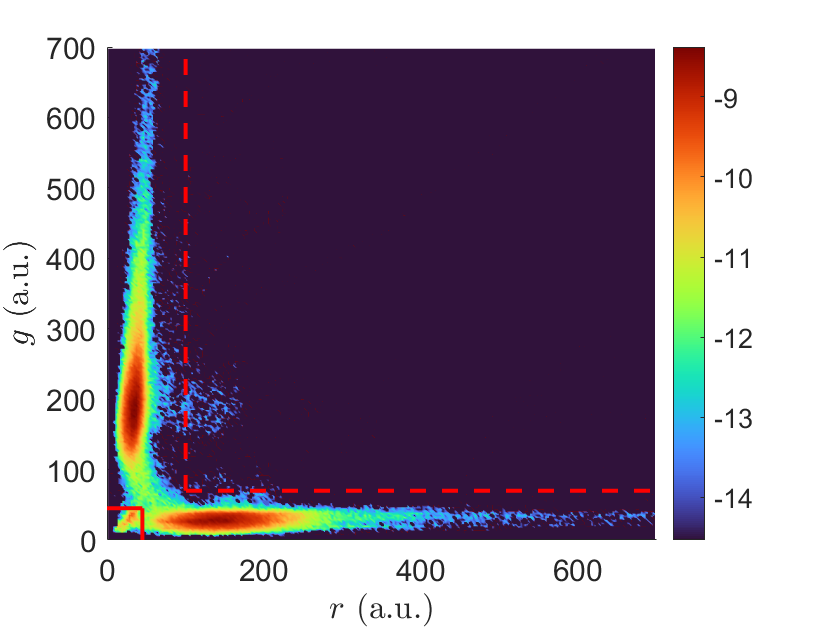}
    \caption{Heatmap of $(r,g)$ values from all cells and all time frames before fluorescence-based filtering. The color represents the logarithm of frequency counts. The solid red lines mark the region \(\{(r,g)\mid r<45,\,g<45\}\), and the dashed red lines indicate the region \(\{(r,g)\mid r>100,\,g>70\}\).
    }
    \label{fig:S2-2}
\end{figure}

\section{Growth Rate Analysis}
\label{gr}

This section describes the procedure used to estimate the time series of growth rates \(\lambda(t)\) from cell length trajectories, and presents the resulting growth rate distributions evaluated under different analytical contexts. The analysis on the growth rates shows the cells are steadily distributed in a fast-growing state over the experiment. According to Ref.~\cite{zhu2023unbalanced}, this implies an effective landscape where the G-state is favored. 

The growth rate of the concerned mother cell is directly estimated from its time series of cell length over generations. After each division, the cell length increases exponentially until the next division event, at which point the length drops substantially (Fig.~\ref{fig:S3-1}(a)). Occasionally, cells may fail to divide, exhibiting growth arrest (Fig.~\ref{fig:S3-1}(b)), or may enter a prolonged growth arrest immediately after division (Fig.~\ref{fig:S3-1}(c)). Divisions and growth arrests are identified for appropriately segment the time series into intervals of exponential growth. For the $i$th segment, the cell length is fitted as  
\begin{equation}
    l_i(t) = l_i(0)\,e^{\lambda_i t}
    \label{eq:exponential_growth}
\end{equation}
The fitting parameter $\lambda_i$ is then assigned to each frame (time point) within this interval. The resulting time series of growth rate are shown in Fig.~\ref{fig:S3-1}(d-f), corresponding to the cell length series in Fig.~\ref{fig:S3-1}(a–c).  

\begin{figure}[htbp]
  \centering
  \begin{overpic}[width=0.85\textwidth]{"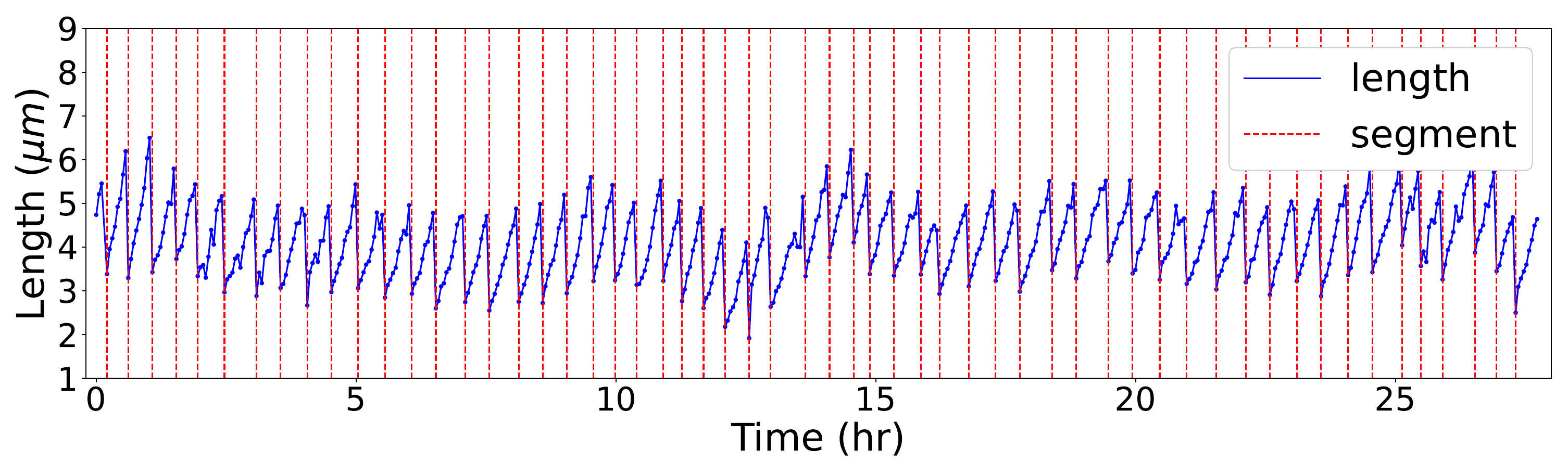"}
    \put(-2,32){\large (a)}
  \end{overpic}
  \begin{overpic}[width=0.85\textwidth]{"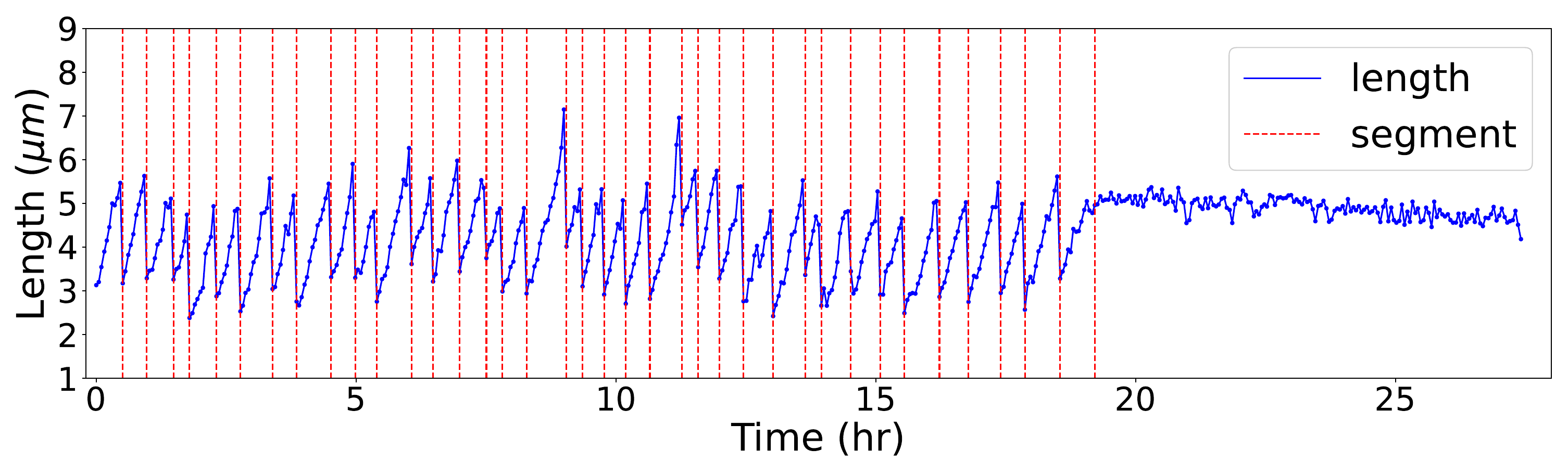"}
    \put(-2,32){\large (b)}   
  \end{overpic}
  \begin{overpic}[width=0.85\textwidth]{"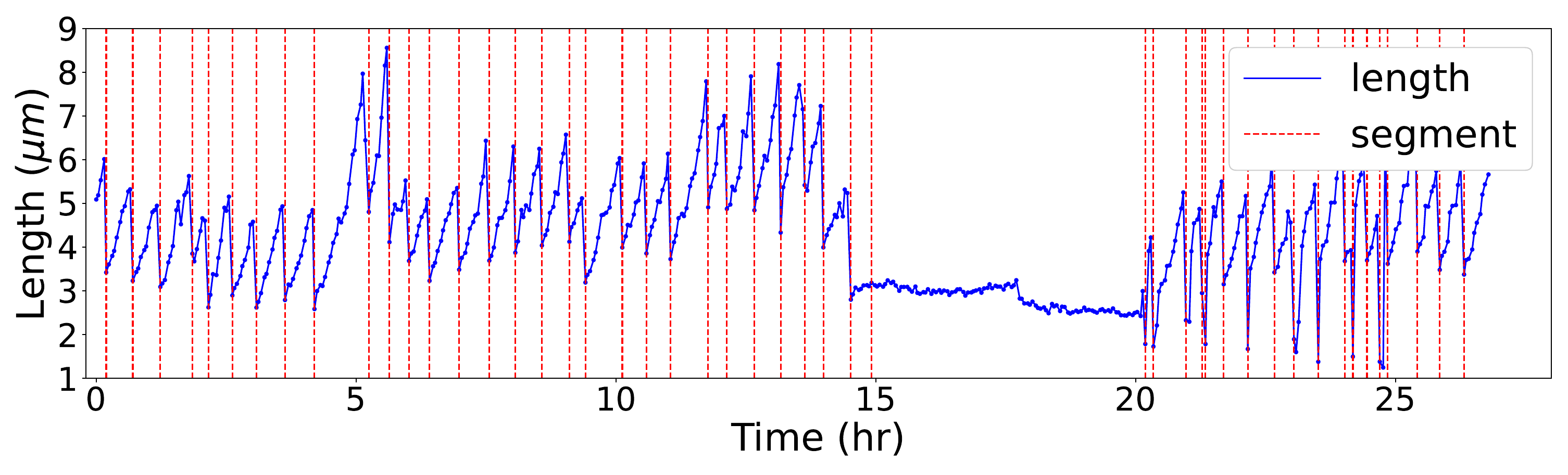"}
    \put(-2,32){\large (c)}  
  \end{overpic}\\[1em]
  \begin{overpic}[width=0.32\textwidth]{"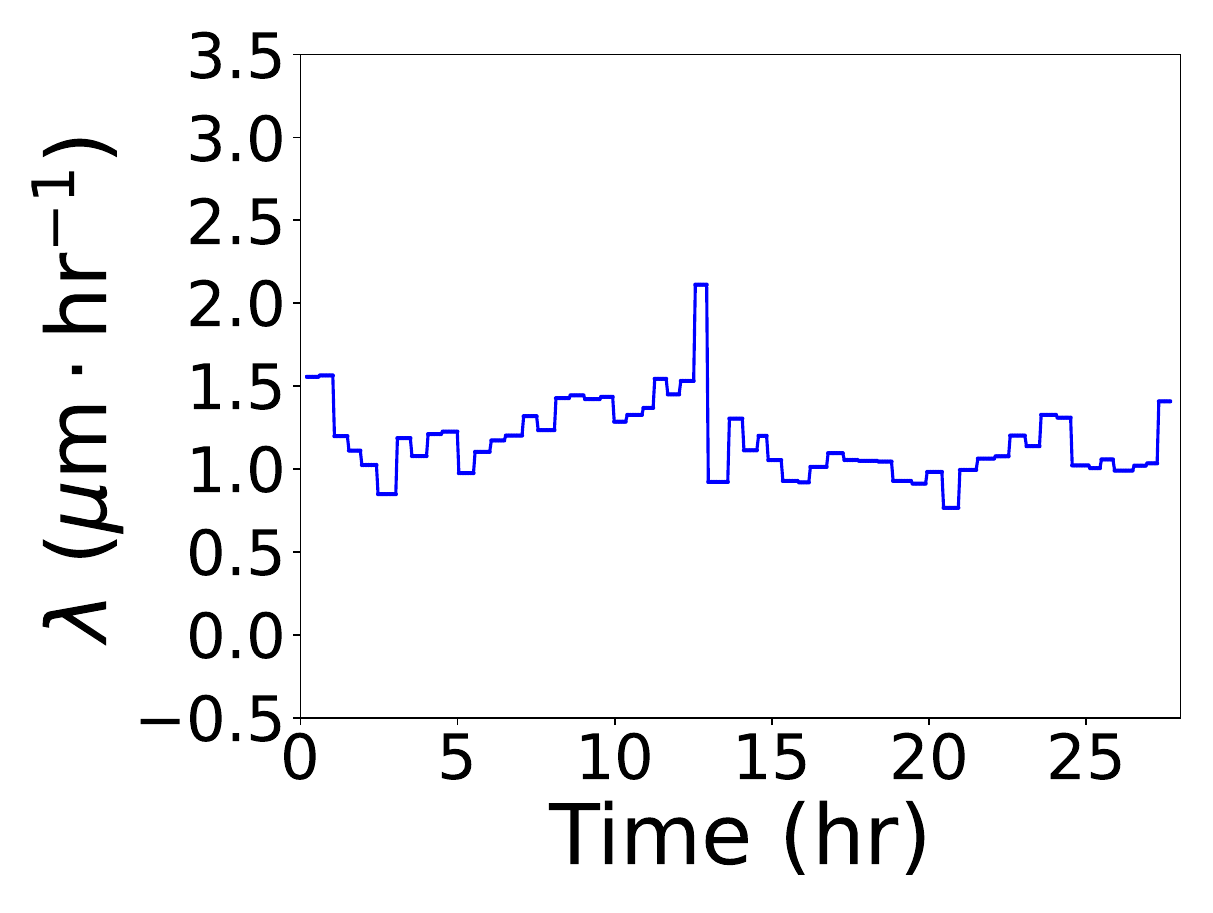"}
    \put(30,63){\large (d)}
  \end{overpic}
  \begin{overpic}[width=0.32\textwidth]{"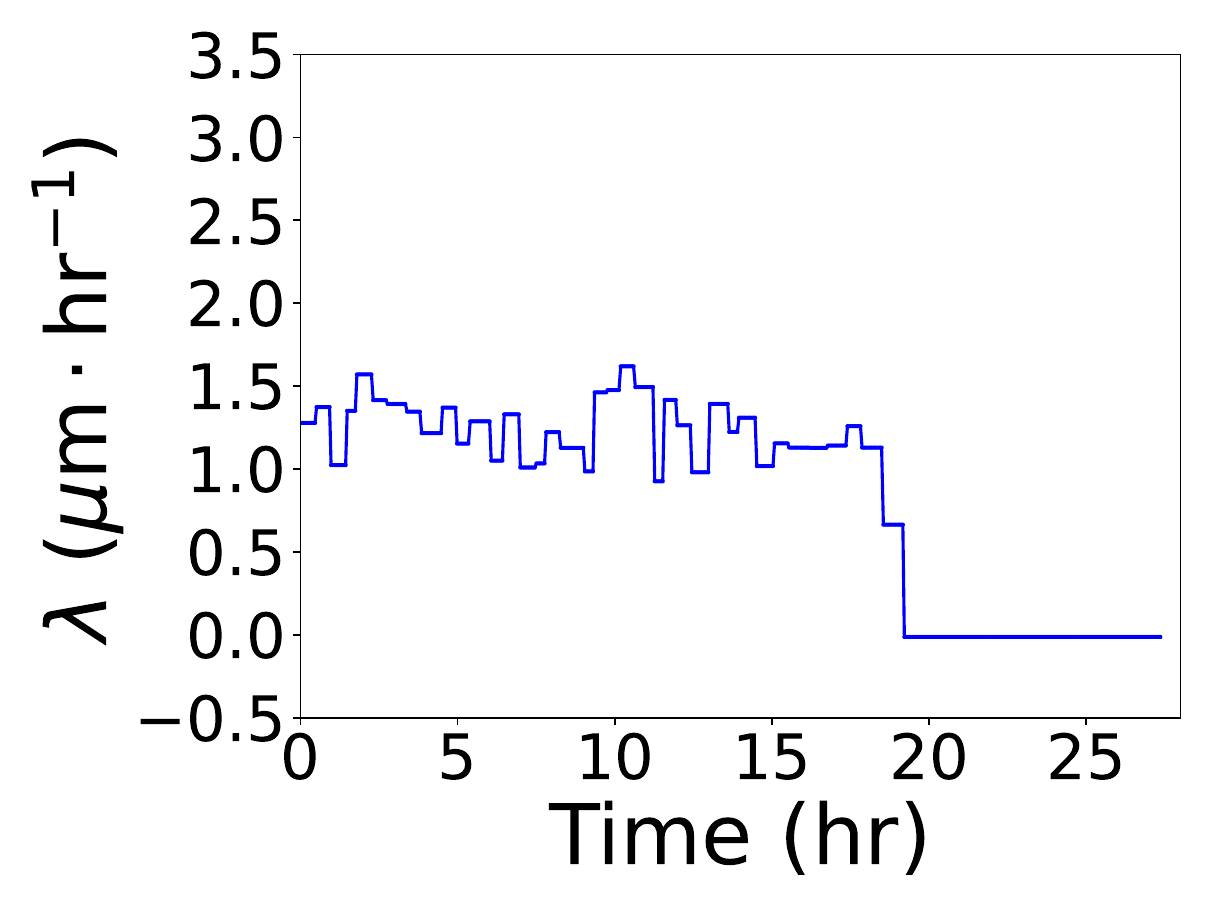"}
    \put(30,63){\large (e)}   
  \end{overpic}
  \begin{overpic}[width=0.32\textwidth]{"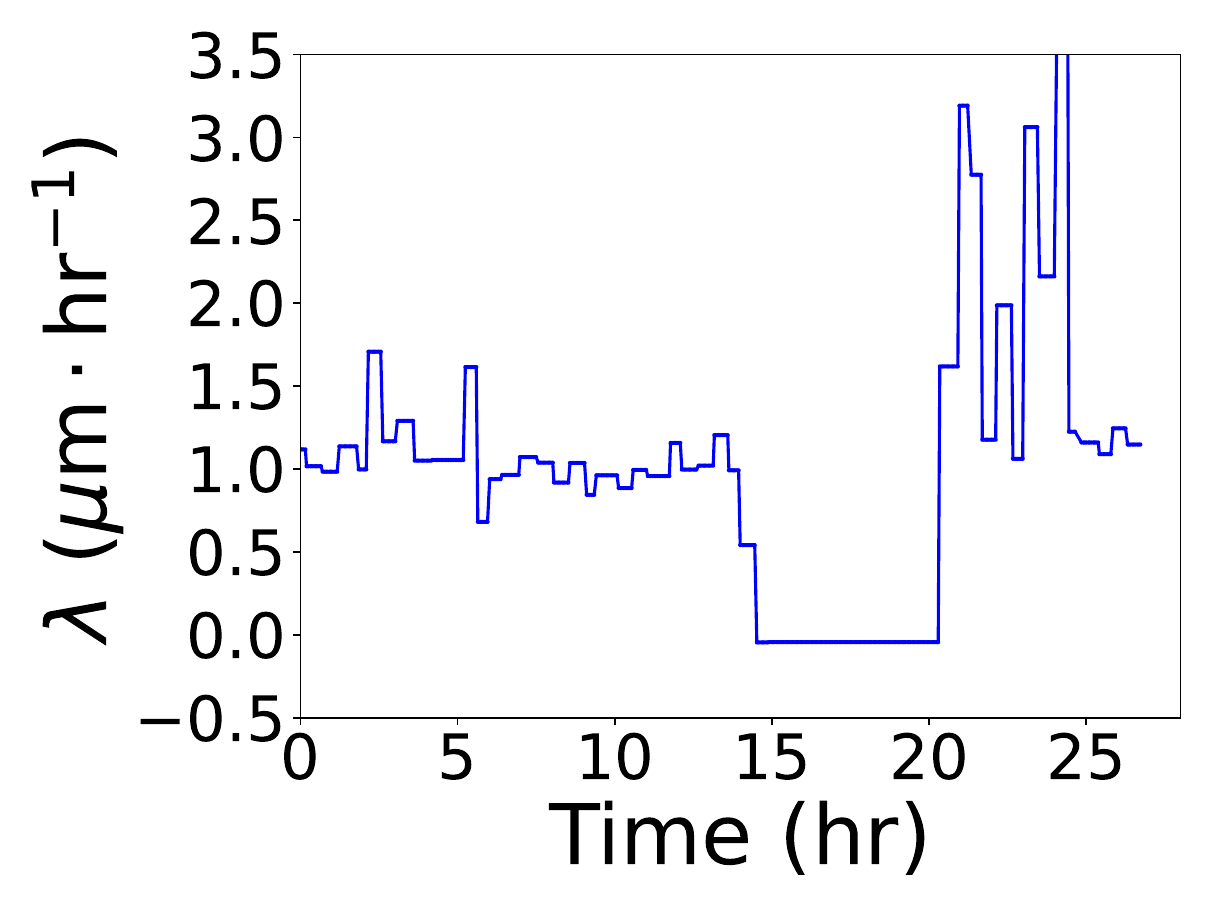"}
    \put(30,63){\large (f)}  
  \end{overpic}
  \caption{The typical time series of cell length (a–c) and the estimated growth rate (d–f) correspond to the same cells shown above. The red dashed vertical lines in (a–c) segment the intervals into continuous elongation phases or growth arrest periods.}
  \label{fig:S3-1}
\end{figure}

From the estimated time series, we calculate growth rate statistics for all cells. The probability density function of $\lambda$ is shown in Fig.~\ref{fig:S3-2}(a). This distribution has a main peak around $1.15$~\(\mu\)m·hr\(^{-1}\). We note a tiny peak near zero indicating growth arrest, contributed by the cells filtered out during data preprocessing in Sec.~\ref{prep}.

The growth rate is a key indicator of cellular physiology state. To address whether the physiology state of the cells are stable over the $27$ hours experiments, we analyzed the growth rate statistics across different experimental periods. The probability density function collapse well, as shown in Fig.~\ref{fig:S3-2}(b), indicating stable physiological state. 

\begin{figure}[H]
  \centering
  \begin{overpic}[width=0.48\textwidth]{"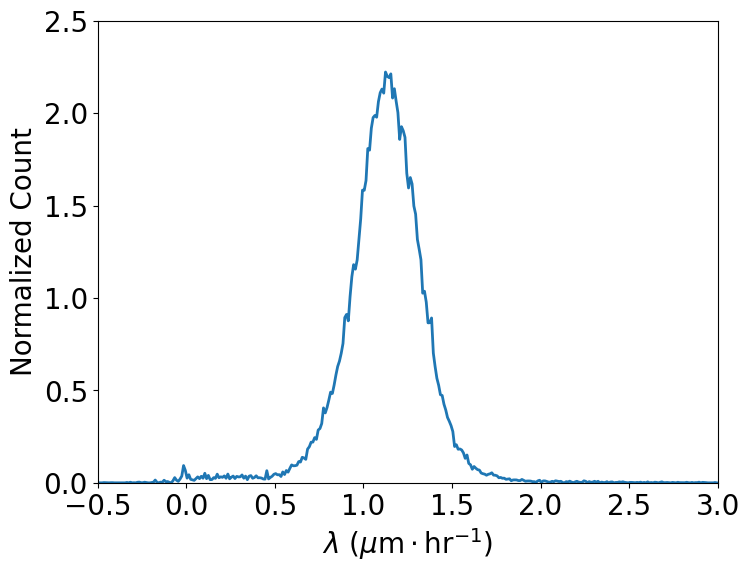"}
    \put(-4,72){\large (a)}
  \end{overpic}
  \begin{overpic}[width=0.48\textwidth]{"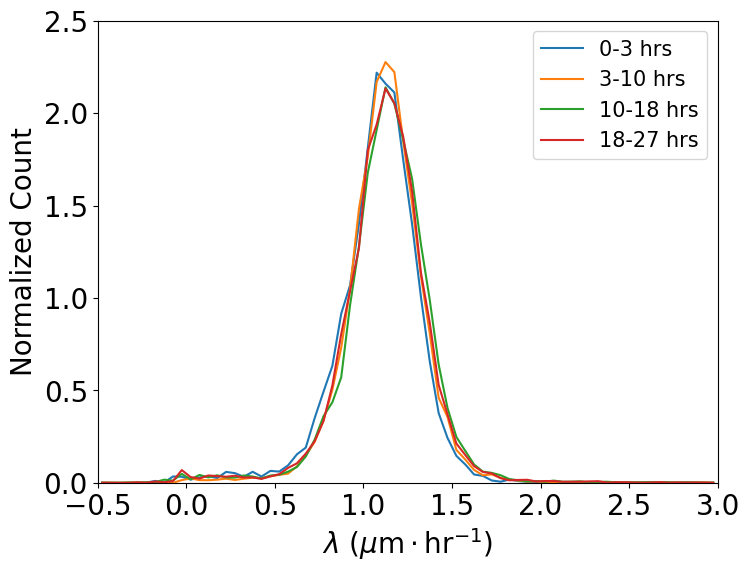"}
    \put(-4,72){\large (b)}   
  \end{overpic}
  \caption{Distribution of growth rate $\lambda$. 
  (a) The distribution of $\lambda$ ($\mu$ m·hr\(^{-1}\)) for all time points from all cells used for further analysis in the main text. 
  (b) The distributions of \(\lambda\) evaluated from various periods of the experiment.}
  \label{fig:S3-2}
\end{figure}

\section{Two-dimensional nonparametric reconstruction and Helmholtz–Hodge decomposition}
\label{sec:2d_recon_hodge}

\subsection{Nonparametric Inference of the Drift Field and Noise in the $(r,g)$ Plane}

We restrict the reconstruction to regions of the $(r,g)$ plane that are well supported by the filtered trajectory ensemble, rather than extrapolating into sparsely sampled areas. Specifically, based on the fluorescence-based filtering criteria described in Sec.~\ref{subsec:fluorescence_filtering}, Eqs.~\eqref{eq:S1}--\eqref{eq:S3}, we perform the reconstruction on $\{(r,g): r\in[0,100],\ g\in[70,500]\}\ \cup\ \{(r,g): r\in[0,400],\ g\in[0,70]\}$. The stochastic dynamics in this region is described by a two-dimensional It\^{o} process,
\begin{equation}
d\mathbf{x}_t=\boldsymbol{\mu}(\mathbf{x}_t)\,dt+\sqrt{2\,\mathbf{D}(\mathbf{x}_t)}\,d\mathbf{W}_t,
\qquad \mathbf{x}_t\equiv (r_t,g_t)^{\mathsf T}.
\label{eq:8-1}
\end{equation}
where $\boldsymbol{\mu}(\mathbf{x})$ is the drift field, and $\mathbf{D}(\mathbf{x})$ is the (multiplicative) noise matrix. $\mathbf{W}_t$ is a standard two-dimensional Wiener process. The trajectories are sampled at a fixed time interval $\Delta t$ (here $\Delta t=0.1548~\mathrm{hour}$), yielding one-step increments $\Delta \mathbf{x}_i=\mathbf{x}_{i+1}-\mathbf{x}_i$.

Following the nonparametric maximum-likelihood reconstruction scheme prompted by Ref.~\cite{ohkubo2011nonparametric} (``simple method''), we infer the drift and noise around a query point $\mathbf{x}_0=(r_0,g_0)^{\mathsf T}$ by approximating the dynamics in its neighborhood with a two-dimensional It\^{o} process with constant coefficients. Under this approximation, the one-step transition in $\mathbb{R}^2$ is Gaussian,
\begin{equation}
p(\Delta \mathbf{x}_i \mid \mathbf{x}_i \approx \mathbf{x}_0; \boldsymbol{\mu},\mathbf{\Sigma})
=\frac{1}{(2\pi)\,|\mathbf{\Sigma}\Delta t|^{1/2}}
\exp\!\left[
-\frac{1}{2}\,(\Delta \mathbf{x}_i-\boldsymbol{\mu}\Delta t)^{\mathsf T}
(\mathbf{\Sigma}\Delta t)^{-1}
(\Delta \mathbf{x}_i-\boldsymbol{\mu}\Delta t)
\right],
\label{eq:8-2}
\end{equation}
where $\boldsymbol{\mu}(\mathbf{x}_0)\in\mathbb{R}^2$ is the local drift and $\mathbf{\Sigma}(\mathbf{x}_0)\in\mathbb{R}^{2\times 2}$ is the local noise matrix, satisfying $\mathbf{\Sigma}(\mathbf{x})=2\,\mathbf{D}(\mathbf{x})$ under the convention of Eq.~(\ref{eq:8-1}).

We use a spatially adaptive anisotropic Gaussian kernel defined through a local metric, because the dataset exhibits anisotropic local variations in $(r,g)$ and the natural local length scales are not well captured by an isotropic kernel in the raw coordinates. Concretely, for each query point $\mathbf{x}_0$ we estimate a local covariance of the point cloud from its $k$ nearest neighbors (Euclidean distance, with $k=200$) among the pooled states ${\mathbf{x}_i}$, following Ref.~\cite{vincent2002manifold}. This choice of $k$ is reasonable, and small variations of $k$ do not qualitatively affect the reconstructed dynamics. Denoting this local covariance by $\mathbf{S}(\mathbf{x}_0)$, we define its Cholesky factor by $\mathbf{S}(\mathbf{x}_0)=\mathbf{L}(\mathbf{x}_0)\mathbf{L}(\mathbf{x}_0)^{\mathsf T}$. Using the locally whitened coordinate $\mathbf{z}=\mathbf{L}(\mathbf{x}_0)^{-1}(\mathbf{x}-\mathbf{x}_0)$, we define a Gaussian kernel function
\begin{equation}
K_h(\mathbf{x}_i-\mathbf{x}_0)\mathrel{:=}\exp\!\left[-\frac{1}{2h^2}\left\|\mathbf{L}(\mathbf{x}_0)^{-1}(\mathbf{x}_i-\mathbf{x}_0)\right\|_2^2\right].
\label{eq:8-3}
\end{equation}
where $h$ is a dimensionless bandwidth in the locally whitened space.

Given the query point $\mathbf{x}_0$, we estimate the local parameters $(\boldsymbol{\mu}(\mathbf{x}_0),\boldsymbol{\Sigma}(\mathbf{x}_0))$ by maximizing the log-likelihood function evaluated on the one-step samples $\mathcal{D}=\{(\mathbf{x}_i,\Delta\mathbf{x}_i)\}_{i=1}^{N}$,
\begin{equation}
\begin{aligned}
\ell\!\left(\boldsymbol{\mu}(\mathbf{x}_0),\boldsymbol{\Sigma}(\mathbf{x}_0)\,\big|\,\mathcal{D}\right)
&=
\sum_{i=1}^{N} K_h(\mathbf{x}_i-\mathbf{x}_0)\,
\log p\!\left(\Delta\mathbf{x}_i\,\big|\,\mathbf{x}_i\approx \mathbf{x}_0;\,\boldsymbol{\mu}(\mathbf{x}_0),\boldsymbol{\Sigma}(\mathbf{x}_0)\right)
\\
&\sim
-\frac{1}{2}\sum_{i=1}^{N} K_h(\mathbf{x}_i-\mathbf{x}_0)
\left[
\log\left|\boldsymbol{\Sigma}(\mathbf{x}_0)\right|
+
\frac{1}{\Delta t}\Big(\Delta\mathbf{x}_i-\boldsymbol{\mu}(\mathbf{x}_0)\Delta t\Big)^{\mathsf T}
\boldsymbol{\Sigma}(\mathbf{x}_0)^{-1}
\Big(\Delta\mathbf{x}_i-\boldsymbol{\mu}(\mathbf{x}_0)\Delta t\Big)
\right].
\end{aligned}
\label{eq:8-4}
\end{equation}
The maximum-likelihood conditions $\nabla_{\boldsymbol{\mu}}\ell=0$ and $\delta \ell/\delta \mathbf{\Sigma}=0$ yield the closed-form estimators
\begin{subequations}\label{eq:8-5-8-6}
\begin{align}
\hat{\boldsymbol{\mu}}(\mathbf{x}_0)
&=
\frac{\sum_i K_h(\mathbf{x}_i-\mathbf{x}_0)\,\Delta \mathbf{x}_i}{\Delta t\sum_i K_h(\mathbf{x}_i-\mathbf{x}_0)},
\label{eq:8-5}
\\
\hat{\mathbf{\Sigma}}(\mathbf{x}_0)
&=
\frac{1}{\Delta t}\,
\frac{\sum_i K_h(\mathbf{x}_i-\mathbf{x}_0)\,(\Delta \mathbf{x}_i-\hat{\boldsymbol{\mu}}(\mathbf{x}_0)\Delta t)(\Delta \mathbf{x}_i-\hat{\boldsymbol{\mu}}(\mathbf{x}_0)\Delta t)^{\mathsf T}}
{\sum_i K_h(\mathbf{x}_i-\mathbf{x}_0)}.
\label{eq:8-6}
\end{align}
\end{subequations}

The bandwidth $h$ controls the bias--variance tradeoff of the reconstruction and is selected by out-of-sample likelihood. We randomly split the one-step dataset $\mathcal{D}$ into a training subset $\mathcal{D}_{\mathrm{tr}}$ (80\% of $\mathcal{D}$) used to obtain $(\hat{\boldsymbol{\mu}}_h,\hat{\mathbf{\Sigma}}_h)$ and a validation subset $\mathcal{D}_{\mathrm{va}}$ (the remaining 20\%) used to score $h$. For each candidate $h$, we compute $(\hat{\boldsymbol{\mu}}_h,\hat{\mathbf{\Sigma}}_h)$ from $\mathcal{D}_{\mathrm{tr}}$ via Eq.~\eqref{eq:8-5-8-6}, and define the negative log-likelihood (NLL) function
\begin{equation}
\overline{\mathrm{NLL}}(h)
=
-\frac{1}{|\mathcal{D}_{\mathrm{va}}|}
\sum_{(\mathbf{x}_i,\Delta\mathbf{x}_i)\in\mathcal{D}_{\mathrm{va}}}
\log p\!\left(\Delta\mathbf{x}_i\,\big|\,\mathbf{x}_i\approx \mathbf{x}_0;\hat{\boldsymbol{\mu}}_h(\mathbf{x}_0),\hat{\mathbf{\Sigma}}_h(\mathbf{x}_0)\right)\Big|_{\mathbf{x}_0=\mathbf{x}_i},
\label{eq:8-7}
\end{equation}
where $p(\cdot)$ is the Gaussian density in Eq.~\eqref{eq:8-2}. We repeat this procedure three times with independently shuffled partitions of $\mathcal{D}$, and obtain three $\overline{\mathrm{NLL}}(h)$ curves shown in Fig.~\ref{fig:S8-1}. The minima consistently occur around $h\simeq 2$, and we therefore choose $h=2$ for the reconstruction.
\begin{figure}[htbp]
    \centering
    \includegraphics[width=0.45\textwidth]{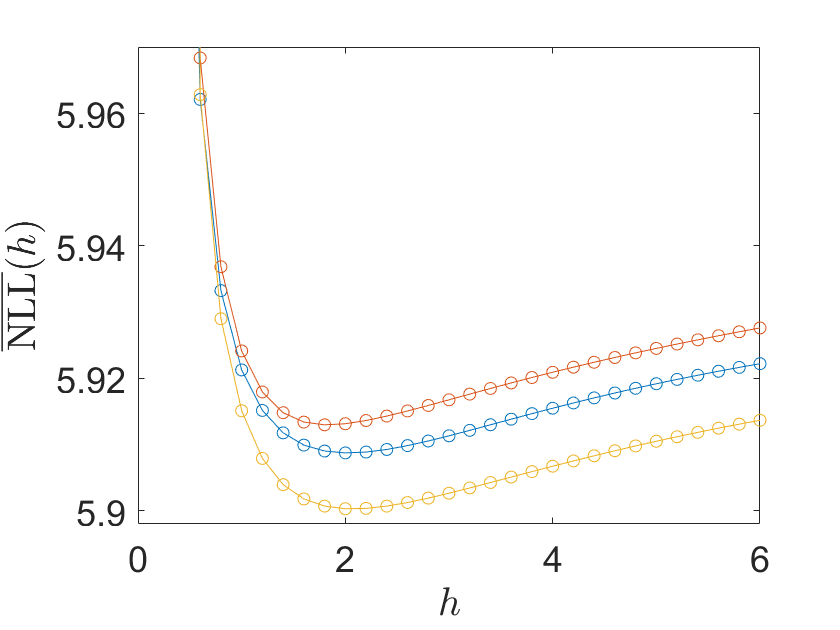}
    \caption{Cross-validation for the kernel bandwidth $h$. The ordinate shows the mean validation negative log-likelihood $\overline{\mathrm{NLL}}(h)$ defined in Eq.~\eqref{eq:8-7}. The three curves correspond to three independently shuffled 80/20 splits of the one-step dataset $\mathcal{D}$. The minima are located near $h\simeq 2$.}
    \label{fig:S8-1}
\end{figure}

To assess the reliability of the local reconstruction, we compute the effective sample size (ESS) associated with the local kernel values. In our setting, for a fixed $\mathbf{x}_0$ we treat $\{K_h(\mathbf{x}_i-\mathbf{x}_0)\}$ as weights and define the normalized weights
\begin{equation}
\bar w_i(\mathbf{x}_0)=\frac{K_h(\mathbf{x}_i-\mathbf{x}_0)}{\sum_j K_h(\mathbf{x}_j-\mathbf{x}_0)},
\label{eq:8-8}
\end{equation}
so that $\{\bar w_i(\mathbf{x}_0)\}$ play the role of normalized importance weights in the self-normalized importance sampling (SNIS) estimator \cite{elvira2022rethinking}. We then define
\begin{equation}
\mathrm{ESS}(\mathbf{x}_0)= \frac{1}{\sum_i \bar w_i(\mathbf{x}_0)^2},
\label{eq:8-9}
\end{equation}
where $\mathrm{ESS}(\mathbf{x}_0)$ provides an empirical diagnostic of whether the local estimates are dominated by a few large normalized weights \cite{elvira2022rethinking}. Fig.~\ref{fig:S8-2} shows that, in our reconstruction region, $\mathrm{ESS}(\mathbf{x}_0)$ is typically large, indicating that the local estimates are not dominated by only a few samples.
\begin{figure}[htbp]
    \centering
    \includegraphics[width=0.45\textwidth]{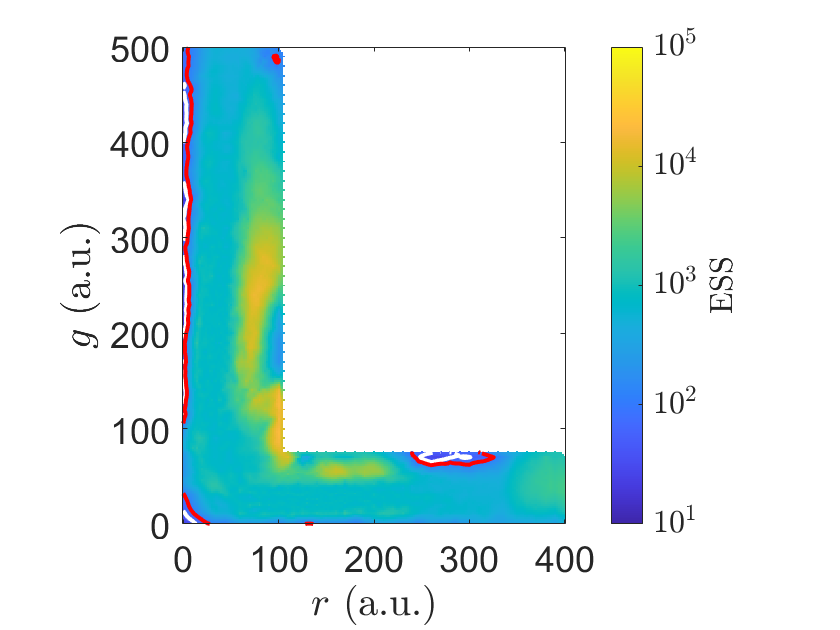}
    \caption{Heatmap of $\mathrm{ESS}(\mathbf{x}_0)$ over the reconstruction region. The white contour indicates $\mathrm{ESS}=50$ and the red contour indicates $\mathrm{ESS}=100$.}
    \label{fig:S8-2}
\end{figure}

The reconstruction yields the drift term $\hat{\boldsymbol{\mu}}(r,g)$ and the noise matrix $\hat{\mathbf{D}}(r,g)$. The drift term is visualized as streamlines in Fig.~\ref{fig:S8-3}(a). The noise term is visualized by the heatmap of $\mathrm{tr}\!\left(\hat{\mathbf{D}}(r,g)\right)$ in Fig.~\ref{fig:S8-3}(b). The strong spatial variation of $\mathrm{tr}\!\left(\hat{\mathbf{D}}(r,g)\right)$ indicates pronounced multiplicative noise. In particular, in the R-state region (the low-$g$ region), larger $r$ is associated with a larger noise magnitude, in agreement with the one-dimensional reconstruction.
\begin{figure}[htbp]
  \centering
  \begin{overpic}[height=0.37\textwidth]{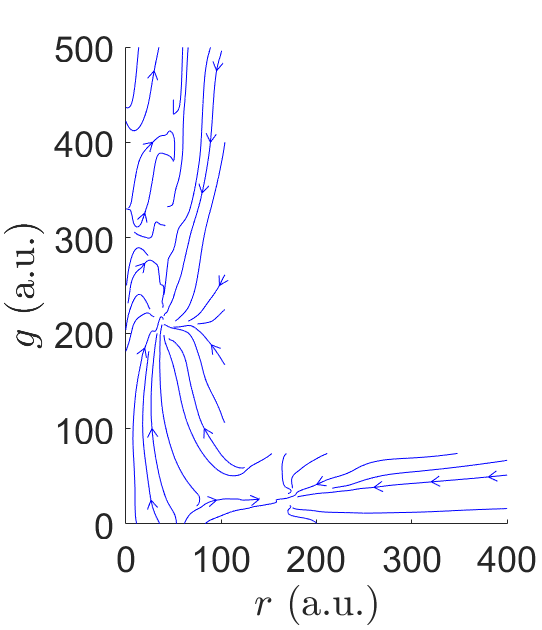}
    \put(-3,95){\large (a)}
  \end{overpic}
  \begin{overpic}[height=0.37\textwidth]{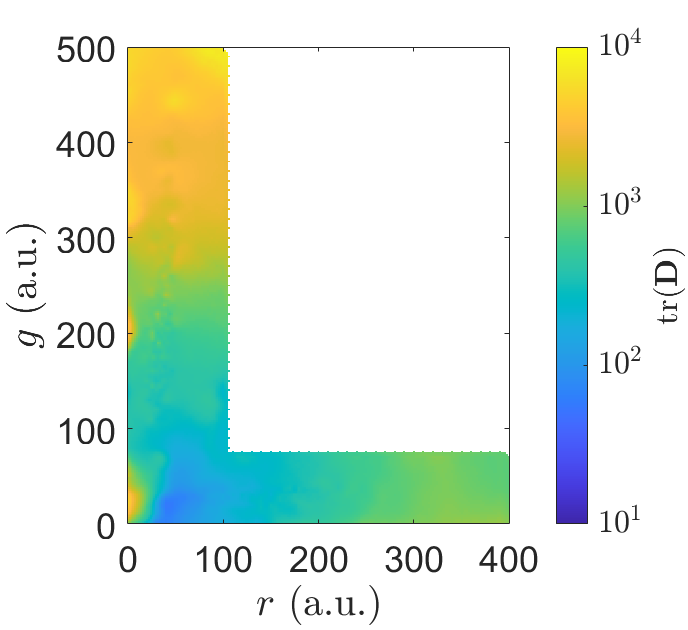}
    \put(-3,86){\large (b)}
  \end{overpic}
  \caption{Reconstruction results in the $(r,g)$ plane.
  (a) Streamlines of the reconstructed drift term $\hat{\boldsymbol{\mu}}(r,g)$.
  (b) Heatmap of the noise magnitude $\mathrm{tr}\!\left(\hat{\mathbf{D}}(r,g)\right)$.}
  \label{fig:S8-3}
\end{figure}

\newpage
\subsection{Helmholtz–Hodge Decomposition of the Reconstructed Drift Field}
\label{subsec:hodge_decomp}

To examine whether the curl component of the reconstructed drift is small enough that the first-passage dynamics can be approximated by a purely gradient flow, we further decompose the drift term $\hat{\boldsymbol{\mu}}(r,g)$ in $\Omega\mathrel{:=}\{(r,g): r\in[20,400],\ g\in[0,66]\}$ as
\begin{equation}
\hat{\boldsymbol{\mu}}(r,g)
=
\nabla \phi(r,g)+\nabla^{\perp}\psi(r,g)+\mathbf{h}(r,g),
\qquad
\nabla^{\perp}\psi \mathrel{:=}
\bigl(\partial_g\psi,\,-\partial_r\psi\bigr)^{\mathsf T},
\label{eq:8-10}
\end{equation}
where $\phi$ determines the gradient component, $\psi$ determines the rotational component, and $\mathbf{h}$ denotes the remainder. Applying the divergence and curl operators to Eq.~\eqref{eq:8-10}, with $\omega \mathrel{:=} \partial_r \hat{\mu}_g-\partial_g \hat{\mu}_r$, yields
\begin{equation}\label{eq:8-11}
\begin{cases}
\Delta \phi = \nabla\cdot \hat{\boldsymbol{\mu}}, \\
\Delta \psi = -\omega,
\end{cases}
\qquad
\text{in } \Omega .
\end{equation}
with boundary conditions
\begin{equation}\label{eq:8-12}
\begin{cases}
\partial_n \phi = \hat{\boldsymbol{\mu}}\cdot \mathbf{n},\quad
\phi(r_\ast,g_\ast)=0,\\
\psi=0,
\end{cases}
\qquad \text{on } \partial\Omega .
\end{equation}
where $\mathbf{n}$ is the outward unit normal and $\partial_n$ denotes the normal derivative.

For the decomposition on $\Omega$, the potentials $\phi$ and $\psi$ are not uniquely determined until the conditions in Eq.~\eqref{eq:8-12} are prescribed. The choice of these conditions is subtle, and different admissible choices generally lead to different decomposition results. Here we impose that the normal flux of the drift is fully assigned to the gradient part. Heuristically, along the axis-aligned boundary of $\Omega$, the gradient part is taken to account for the component that crosses the boundary, whereas the rotational part is tangent to the boundary. This is achieved by requiring $\psi$ to be constant on $\partial\Omega$, and we therefore set $\psi=0$ on $\partial\Omega$. Altogether, this leads to the boundary conditions in Eq.~\eqref{eq:8-12}.

Because $\hat{\boldsymbol{\mu}}$ is reconstructed on a discrete set of query points, we solve Eq.~\eqref{eq:8-11} subject to the boundary conditions \eqref{eq:8-12} on a uniform Marker-and-Cell (MAC) grid, using the standard second-order finite-difference discretization of the Laplacian. Under the boundary conditions \eqref{eq:8-12}, Eq.~\eqref{eq:8-10} on the simply connected domain $\Omega$ implies $\mathbf{h}\equiv 0$. In the discrete computation, we verify that $\mathbf{h}$ remains very small (see Fig.~\ref{fig:S8-5}(b)), consistent with $\mathbf{h}\equiv 0$.

The decomposition results are shown in Figs.~\ref{fig:S8-4} and \ref{fig:S8-5}. We plot $U(r,g)=-\phi(r,g)$ in Fig.~\ref{fig:S8-4}, where $U$ plays the role of the potential associated with the gradient component. Fig.~\ref{fig:S8-5}(a) shows the magnitudes of the gradient and rotational components. The rotational component is much smaller than the gradient component over most of $\Omega$, as can be seen directly from the comparison of their magnitudes. Accordingly, we could regard the gradient part as an adequate approximation to the drift term for the two-dimensional first-passage analysis.
\begin{figure}[htbp]
  \centering
  \begin{overpic}[width=0.32\textwidth]{"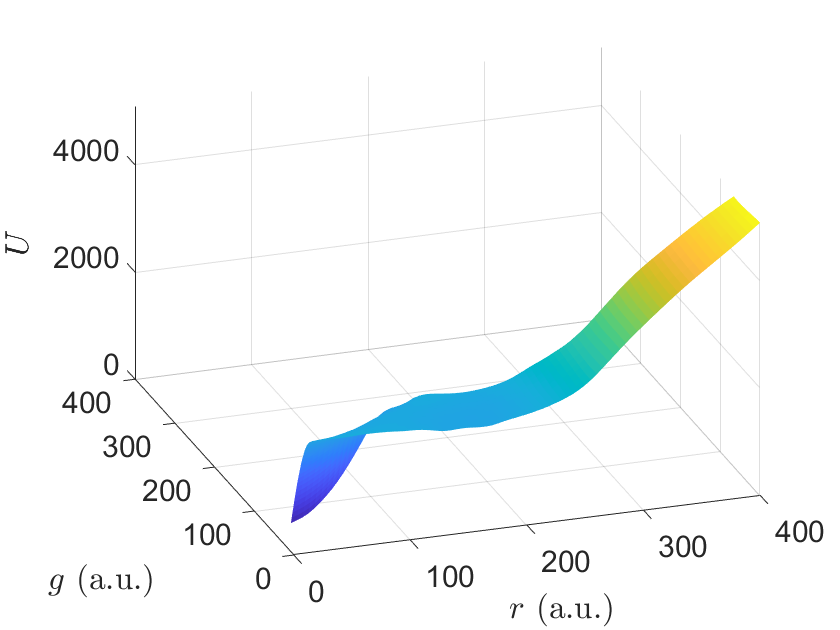"}
    \put(-5,70){\large (a)}
  \end{overpic}
  \begin{overpic}[width=0.32\textwidth]{"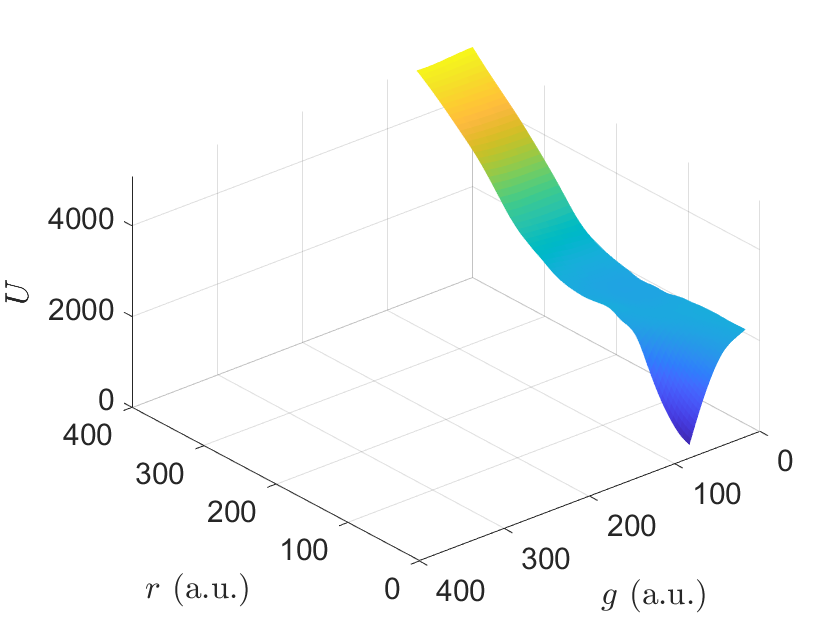"}
    \put(-5,70){\large (b)}
  \end{overpic}
  \begin{overpic}[width=0.32\textwidth]{"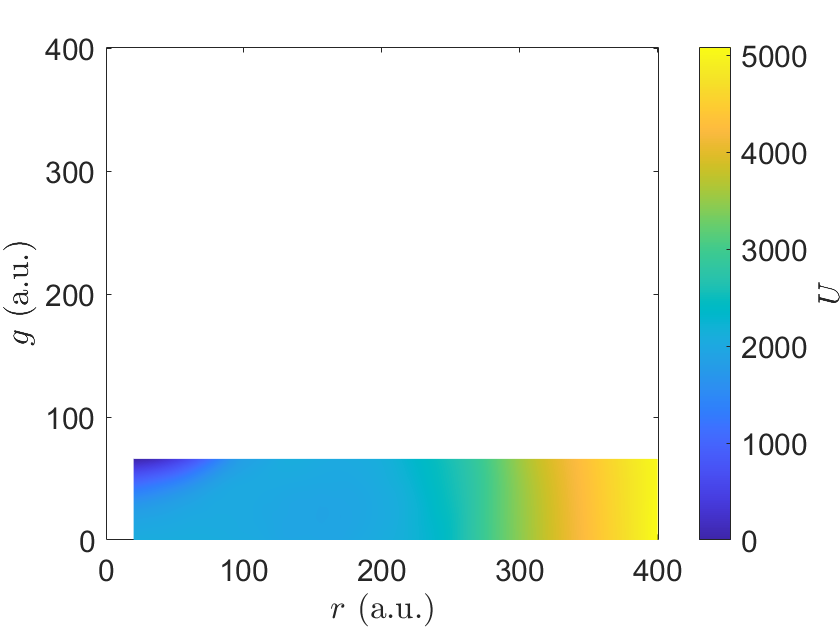"}
    \put(-5,70){\large (c)}
  \end{overpic}
  \caption{(a,b) Three-dimensional views of $U(r,g)$ from two perspectives. (c) Heatmap of $U(r,g)$.}
  \label{fig:S8-4}
\end{figure}

\begin{figure}
  \centering
  \includegraphics[width=0.8\textwidth]{"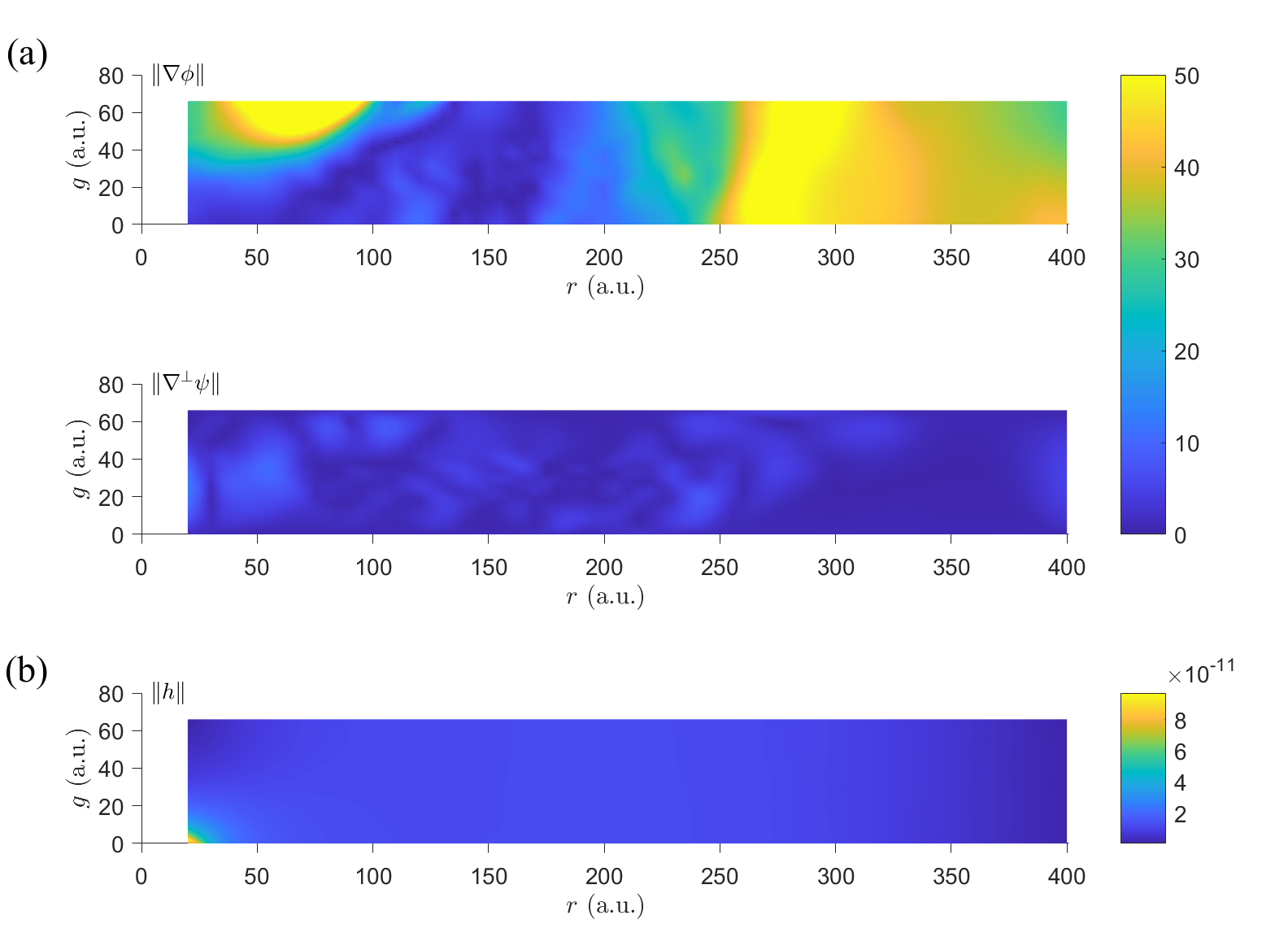"}
  \caption{(a) Heatmaps of $\|\nabla\phi\|$ (top) and $\|\nabla^\perp\psi\|$ (bottom). (b) Heatmap of $\|\mathbf{h}\|$.}
  \label{fig:S8-5}
\end{figure}

\section{Selection of the absorbing boundary from the reconstructed two-dimensional dynamics}
\label{sec:2d_absorbing_boundary}

In this section, we determine the effective absorbing boundary for the first-passage analysis based on the reconstructed drift field and the potential $U(r,g)=-\phi(r,g)$ obtained in Sec.~\ref{subsec:hodge_decomp}. The saddle region of $U(r,g)$ is relatively flat and extended, as shown in Fig.~\ref{fig:S9-1}(b,c), so placing the absorbing boundary further toward the other basin would also include additional post-saddle dynamics. Since that part is not the focus of the present analysis, we instead choose the absorbing boundary in the saddle region so as to isolate the barrier-climbing part of the escape process. For each fixed $g$ in the saddle region, we locate the local maximum of $U$ along the $r$ direction and connect these points into a ridge line, shown as the black curve in Fig.~\ref{fig:S9-1}(a--e). Based on this ridge line, we obtain a linear approximation, $g=kr+b$ with $k=0.7737$ and $b=-42.32$, shown as the red line in Fig.~\ref{fig:S9-1}(a--e), and take this fitted line as the absorbing boundary for the two-dimensional first-passage analysis.

It should be noted from Fig.~\ref{fig:S9-1} that this red line provides a good characterization of the saddle ridge in both $U(r,g)$ (Fig.~\ref{fig:S9-1}(a--c)) and the reconstructed drift term $\hat{\boldsymbol{\mu}}(r,g)$ (Fig.~\ref{fig:S9-1}(d)). By contrast, in the empirical distribution obtained by pooling all trajectories and all time points (Fig.~\ref{fig:S9-1}(e)), the same line appears to cut through the interior of the $r$-state distribution rather than trace its outer boundary. We believe that this is because the resulting distribution is still not a steady-state one, since ergodicity is not achieved yet. This further shows that, in both one and two dimensions, it is important to infer the drift and diffusion terms from the first and second moments of $\Delta r$ (Eq.~(\ref{recf}--\ref{recd}) in the main text) or $\Delta \mathbf{x}$ (Eq.~\eqref{eq:8-5-8-6} for the two-dimensional case), rather than infer the dynamics directly from a pooled empirical occupancy distribution, such as an ensemble snapshot distribution.
\begin{figure}[htbp]
  \centering
  \begin{overpic}[width=0.32\textwidth]{"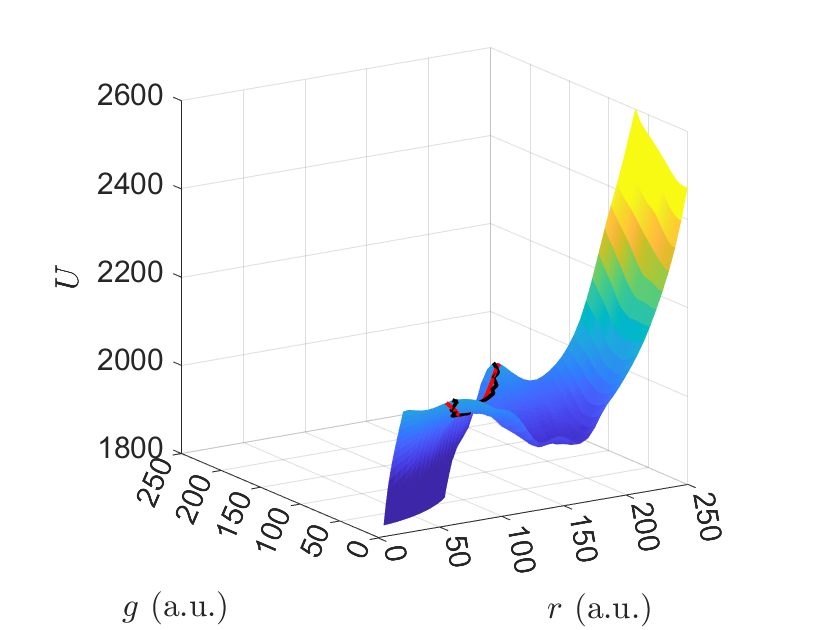"}
    \put(-1,65){\large (a)}
  \end{overpic}
  \begin{overpic}[width=0.32\textwidth]{"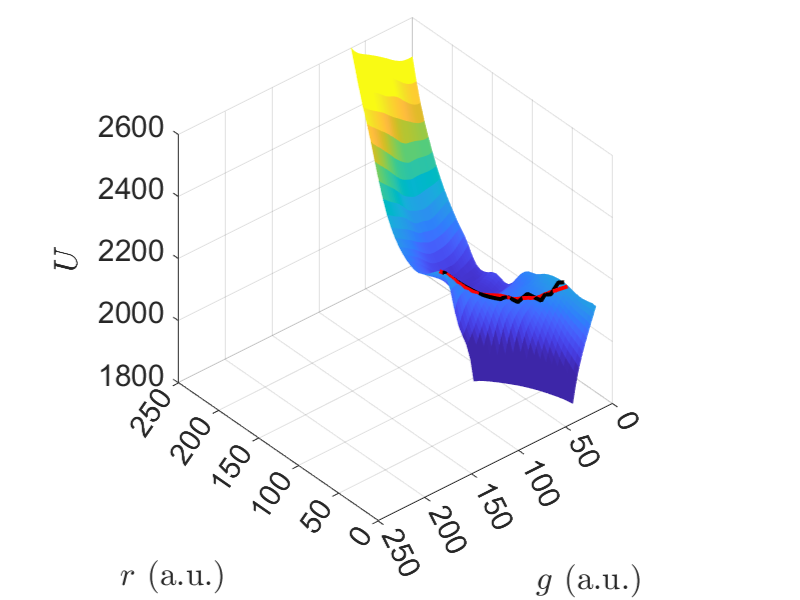"}
    \put(-1,65){\large (b)}
  \end{overpic}
  \begin{overpic}[width=0.32\textwidth]{"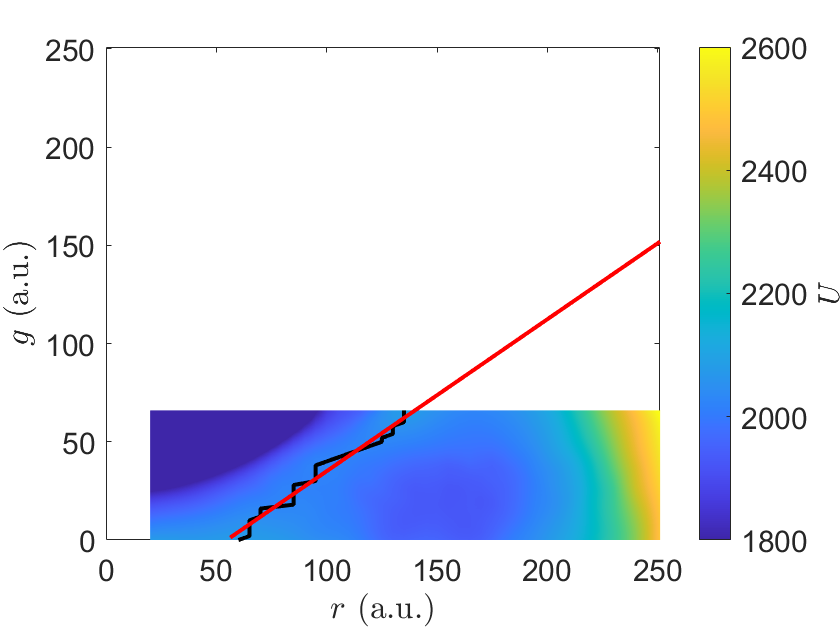"}
    \put(-5,65){\large (c)}
  \end{overpic}
  \begin{overpic}[width=0.40\textwidth]{"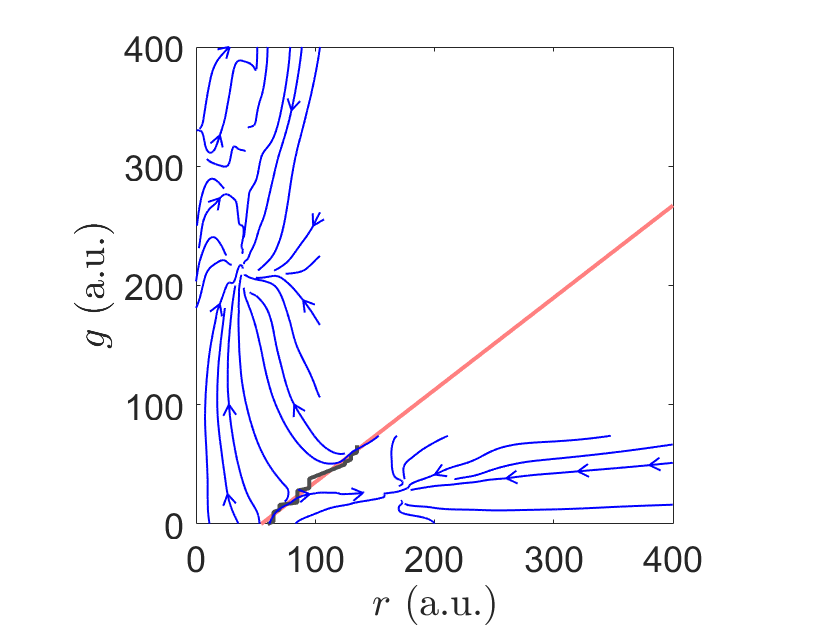"}
    \put(4,68){\large (d)}
  \end{overpic}
  \begin{overpic}[width=0.40\textwidth]{"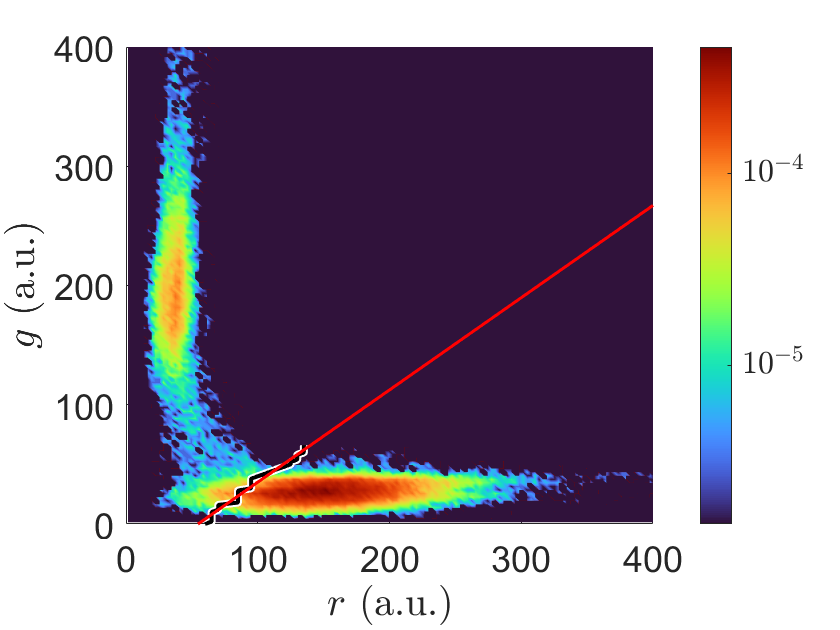"}
    \put(-4,68){\large (e)}
  \end{overpic}
  \caption{Selection of the two-dimensional absorbing boundary. (a,b) Three-dimensional views of $U(r,g)$ from two perspectives, and (c) its heatmap. The black curve marks the ridge line identified in the saddle region, and the red line is its linear fit, which is taken as the absorbing boundary. (d) Streamlines of the reconstructed drift field $\hat{\boldsymbol{\mu}}(r,g)$, and (e) the empirical occupancy distribution in the $(r,g)$ plane, both overlaid with the same ridge line and absorbing boundary.}
  \label{fig:S9-1}
\end{figure}

For the effective one-dimensional dynamics in the main text, we choose the absorbing boundary based on the first-passage events to the two-dimensional absorbing boundary. The distribution of the corresponding $r$ coordinate is shown in Fig.~\ref{fig:S9-2}. Its mean is $84$ (red dashed line in Fig.~\ref{fig:S9-2}); therefore, we choose $r_a=84$ as the absorbing boundary in the one-dimensional analysis. This value is also very close to the barrier position $r_b=86$ of the reconstructed one-dimensional landscape in the main text, further supporting the consistency between the two-dimensional reconstruction and the effective one-dimensional reduction.
\begin{figure}[H]
  \centering
  \includegraphics[width=0.5\textwidth]{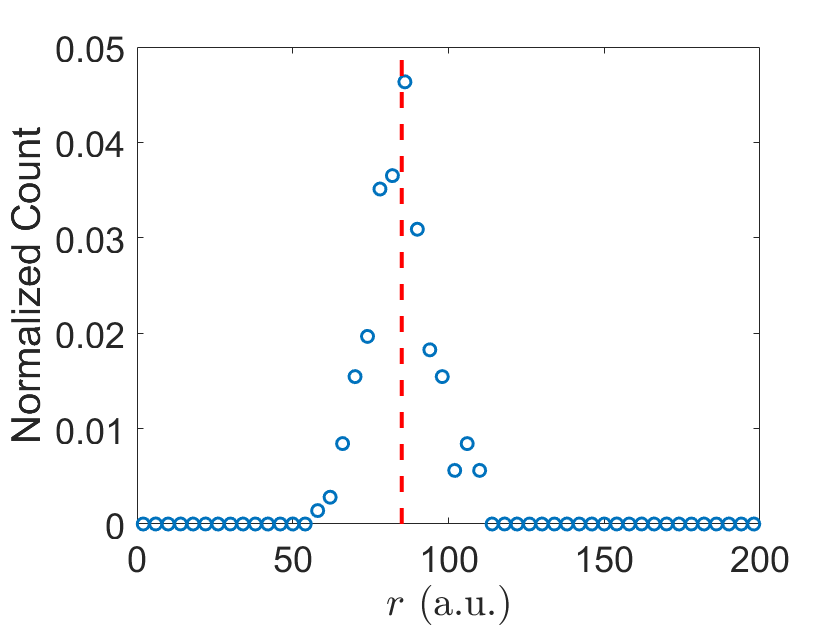}
  \caption{The distribution of the position ($r$) of the first-passage event to the two-dimensional absorbing boundary. The red dashed line marks the mean value ($r=84$), which is chosen as the absorbing boundary for the effective one-dimensional dynamics. The variance of the distribution is $110$.}
  \label{fig:S9-2}
\end{figure}

\newpage
\section{First passage Statistics of Single-Cell Trajectories}
\label{fp}

In this section, we provide further details on the first-passage analysis of the experimental data. It should be noted that, at $t=0$, 477 of the 1,007 cells remain in the R-state region, and all first-passage statistics in this section are computed from this subset.

The first passage time (FPT), denoted as $\tau(r_0, g_0)$, is the time required for a particle to reach the absorbing boundary from the initial state $(r_0 = r(t = 0), g_0 = g(t = 0))$. Figure~\ref{fig:S5-1}(a) shows the FPT for cells initialized at specific $(r_0, g_0)$ values, as observed in the experiment, where each symbol represents a single cell and the color indicates the first passage time to the absorbing boundary, marked by the red straight line in Fig.~\ref{fig:S5-1} and described in Sec.~\ref{sec:2d_absorbing_boundary}. Fig.~\ref{fig:S5-1}(b) shows the corresponding local finite-time mean FPT (MFPT) obtained by bin averaging over the initial-state plane. The FPT is found to depend significantly on the initial state $(r_0, g_0)$, with cells initialized closer to the boundary exhibiting shorter transition times. This intuitive finding leads to the key discovery of this study—namely, that the small-noise limit is not applicable in the current case.

\begin{figure}[htbp]
  \centering
  \begin{overpic}[width=0.55\textwidth]{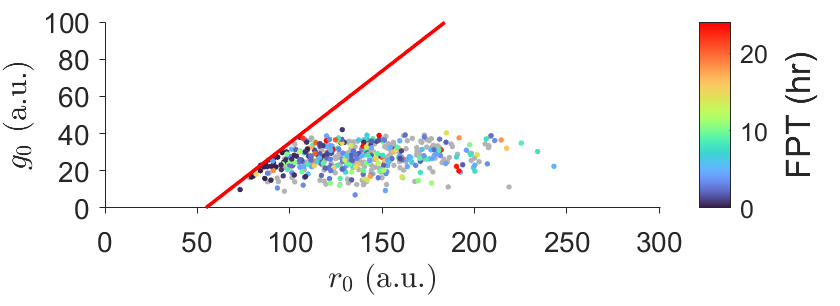}
    \put(-5,34){\large (a)}
  \end{overpic}
  \begin{overpic}[width=0.55\textwidth]{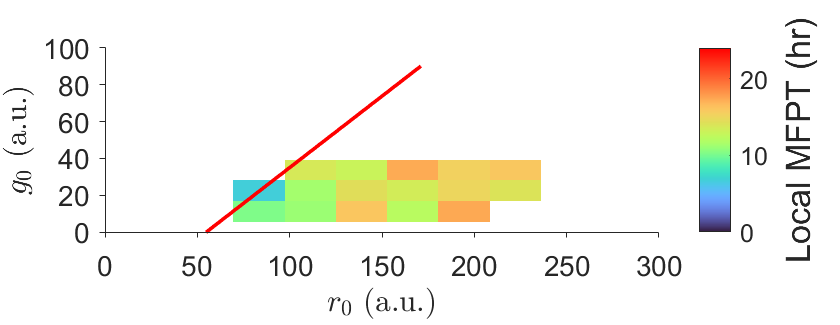}
    \put(-5,34){\large (b)}
  \end{overpic}
  \caption{First-passage statistics as a function of the initial state. (a) FPT of individual single-cell trajectories plotted against the initial state $(r_0,g_0)$, where $r_0=r(t=0)$ and $g_0=g(t=0)$. Each symbol represents one cell, and the symbol color indicates the FPT to the absorbing boundary. Gray symbols denote cells that did not reach the absorbing boundary within the experimental period. (b) Local finite-time MFPT obtained by bin averaging over the initial-state plane. The red line indicates the absorbing boundary used in the analysis.}
  \label{fig:S5-1}
\end{figure}

The full information of the FPT can be characterized by the probability density function $F(\tau \vert r_0, g_0)$, also known as the FPT distribution. Due to limited statistics, the FPT distribution is typically only roughly estimated from experimental data. An alternative measure is the MFPT, defined as
\begin{equation}
\left\langle \tau \right\rangle_{(r_0,g_0)} = \int_0^{\infty} d\tau \, \tau F(\tau \vert r_0, g_0).
\end{equation}
However, achieving the true long-time regime in experiments with finite observation periods is difficult. As a result, direct statistics of the first passage time yield a truncated MFPT
\begin{equation}
\tau'_{(r_0,g_0)} = \int_0^{t_c} d\tau \, \tau F(\tau \vert r_0, g_0),
\end{equation}
where $t_c$ is the cutoff time. This truncation introduces a systematic bias, since $\tau' < \left\langle \tau \right\rangle$.
A more accurate approach is to use statistics based on the survival probability $S(t \vert r_0, g_0)$, defined as the probability that a particle has not yet reached the absorbing boundary by time $t$. The FPT distribution can then be evaluated as
\begin{equation}
F(\tau \vert r_0, g_0) = -\left. \frac{\partial}{\partial t} S(t \vert r_0, g_0) \right|_{t=\tau}.
\end{equation}
Thus, the survival probability also contains the full information of the FPT. As a cumulative distribution function, $S(t)$ behaves more robustly than the probability density function $F(\tau)$, particularly in cases with limited statistics conditioned on the initial state $(r_0, g_0)$. 

From the experimental data, the survival probability $S(t)$ can be obtained by calculating the fraction of cells that have not yet reached the absorbing boundary at time $t$, i.e., $S(t)=N_{\mathrm{survived}}(t)/N(t=0)$. Figure~\ref{fig:S5-2} shows the survival probability relative to the absorbing boundary. Over the whole experiment period, $62.05\%$ of the cells reached the boundary. Fig.~\ref{fig:S5-2}(b,c) shows that the survival curve exhibits non-exponential behavior, indicating that the transition process is not characterized by a single exponential timescale.
\begin{figure}[htbp]
\centering
\begin{overpic}[width=0.32\textwidth]{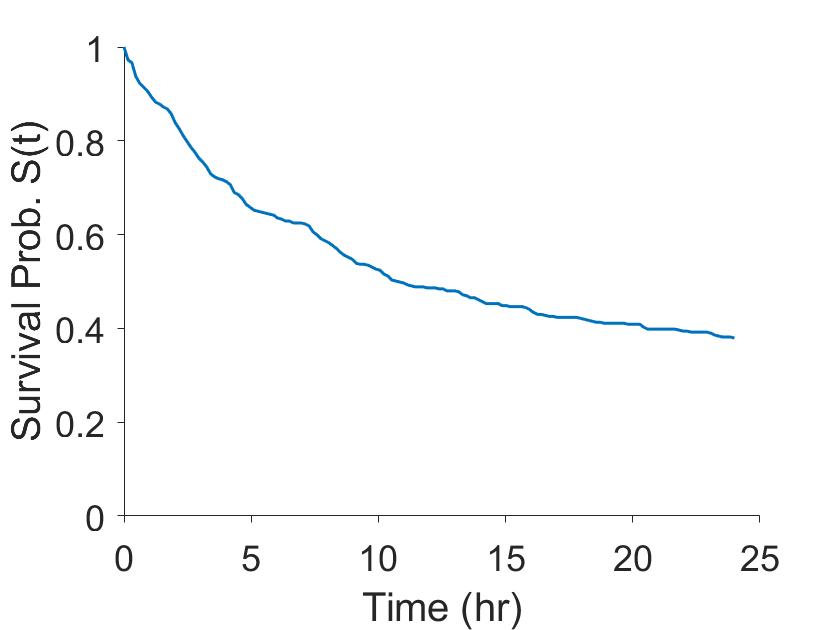}
  \put(-4,68){\large (a)}
\end{overpic}
\begin{overpic}[width=0.32\textwidth]{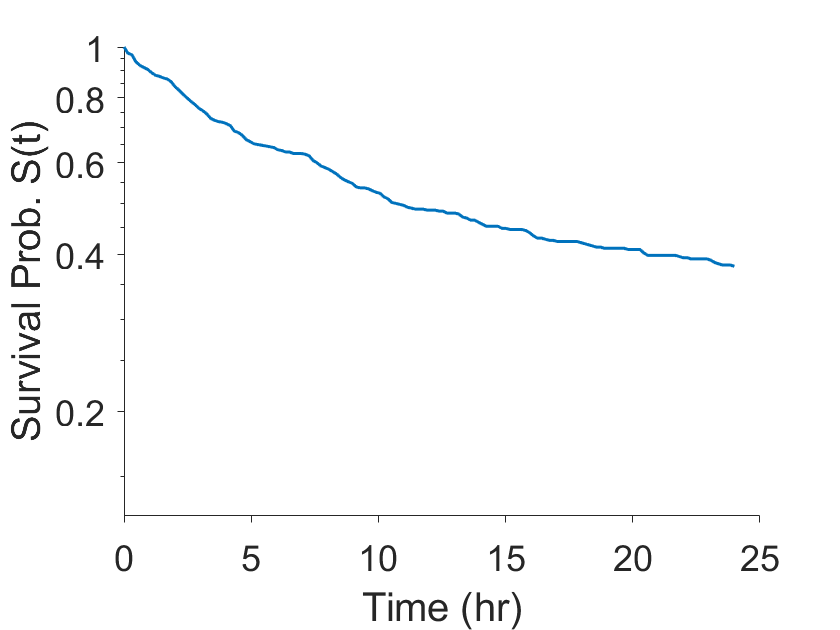}
  \put(-4,68){\large (b)}
\end{overpic}
\begin{overpic}[width=0.32\textwidth]{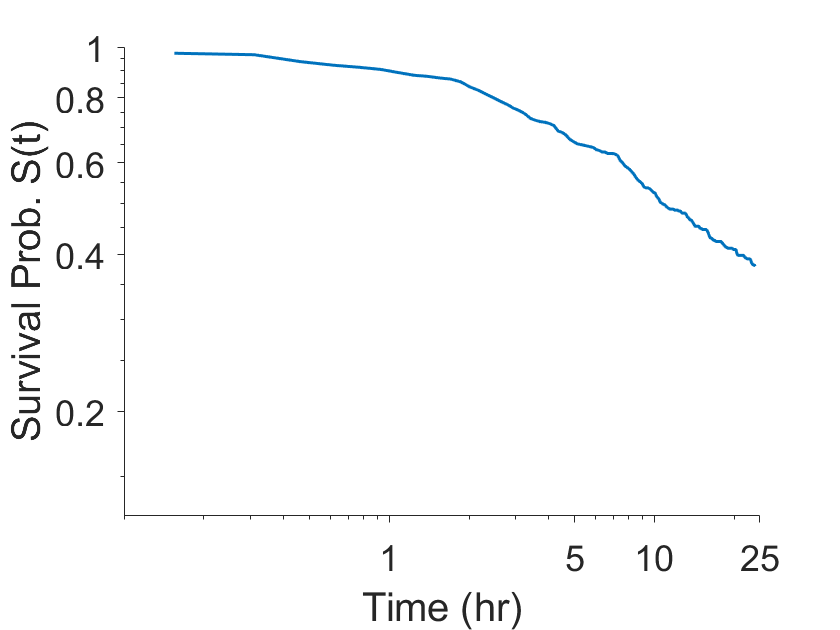}
  \put(-4,68){\large (c)}
\end{overpic}
\caption{Survival probability $S(t)$ for all cells initialized in the $r$-state region, shown on (a) linear, (b) semi-logarithmic, and (c) log-log scales. By the end of the experiment, about $62.05\%$ of the cells have reached the absorbing boundary.}
\label{fig:S5-2}
\end{figure}

To examine the dependence on the initial state, survival statistics were analyzed for cells initialized within specific regions of the $(r, g)$ plane. Figure~\ref{fig:S5-3}(a) identifies 5 such initial regions. The survival probabilities differ significantly depending on the initial state, as shown in Fig.~\ref{fig:S5-3}(b). These results confirm the intuitive observation from Fig.~\ref{fig:S5-1} that cells initialized closer to the boundary require shorter transition times.
\begin{figure}[htbp]
  \centering
  \begin{overpic}[width=0.5\textwidth]{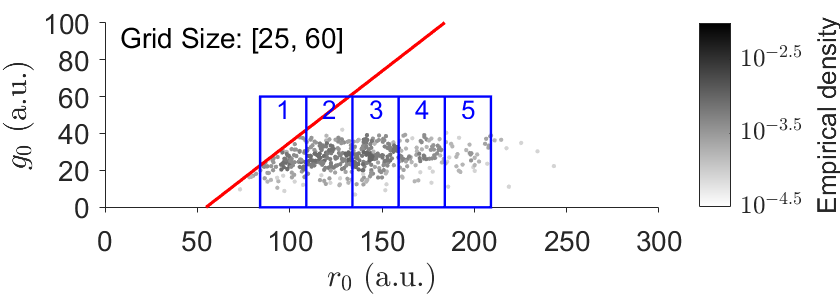}
    \put(-5,34){\large (a)}
  \end{overpic}
  \begin{overpic}[width=0.5\textwidth]{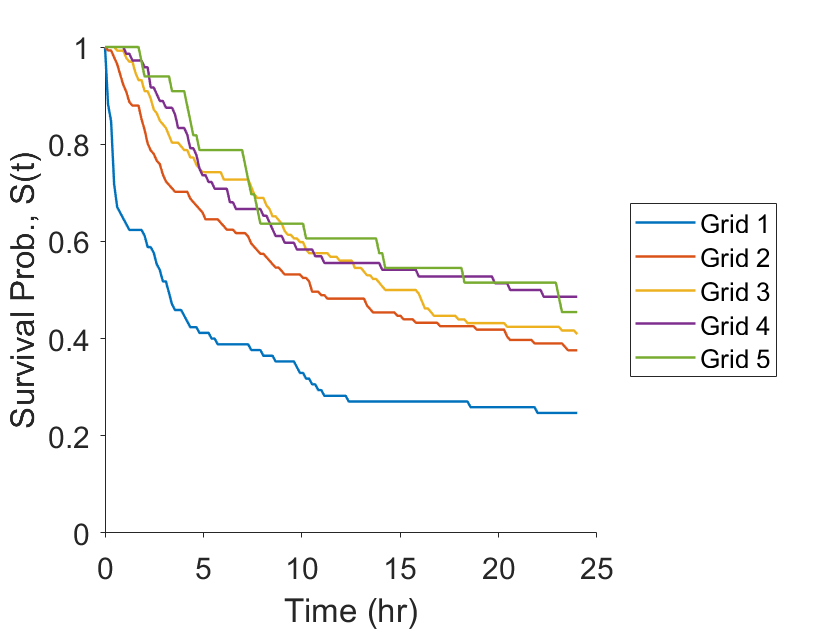}
    \put(-5,70){\large (b)}  
  \end{overpic}
  \caption{(a) Empirical density of the initial states in the $(r, g)$ plane. The red line indicates the absorbing boundary. Five specific initial regions are marked in blue. The $g$-dimension is undivided, with $0<g<60$. The $n$th region in the $r$-dimension spans from $r_0+(n-1)\,dr$ to $r_0+n\,dr$. (b) Survival probabilities $S_n(t)$ for cells initialized in the five regions marked in (a).}
  \label{fig:S5-3}
\end{figure}

Figure~\ref{fig:S5-4} further shows that the finite-time MFPT depends clearly on $r_0$. This trend is consistent with the simulated MFPT shown in Fig.~\ref{fig4} of the main text, although some fluctuations remain because of the limited number of experimentally observed transition events.
\begin{figure}[htbp]
\centering
\includegraphics[width=0.55\textwidth]{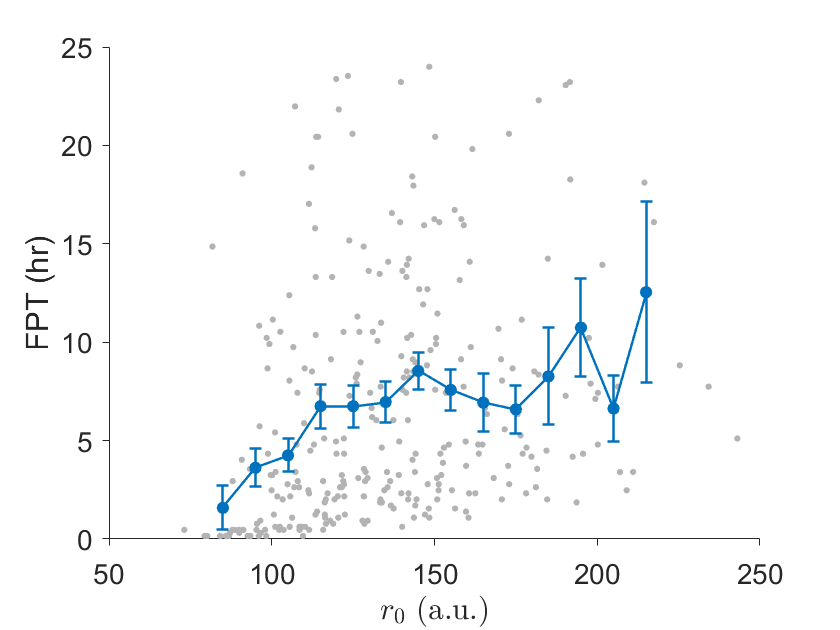}
\caption{FPT vs $r_0$ for cells that reached the absorbing boundary within the experimental period. Gray symbols represent individual cells. Blue points show the bin-averaged MFPT.}
\label{fig:S5-4}
\end{figure}

\section{Analysis on the Reconstruction of U(r) and D(r)}
\label{rec}

This section provides the details on reconstruction of the effective landscape and the noise strength from single cell trajectories, especially checking the robustness of the reconstructed dynamics under various spatial resolutions, temporal resolutions and in various experiment periods. 

As shown in Eq.~(\ref{fpe}) in the main text, the stochastic dynamics are reconstructed in the form of Fokker-Planck equation as
\begin{equation}
\label{eq:fpe}
\frac{\partial P(r,t)}{\partial t}=-\frac{\partial}{\partial r}f(r)P(r,t)+\frac{\partial^2}{\partial r^2}D(r)P(r,t). 
\end{equation}
The drift force $f(r)$ and the noise strength $D(r)$ are estimated from the single cell trajectories following Eqs.~(\ref{recf}--\ref{recd}) in the main text as
\begin{eqnarray}
f(r) & = & \frac{1}{\Delta t}\left<r(r+\Delta t)-r(t)\right>\vert_{r(t)=r},\\
D(r) & = & \frac{1}{2\Delta t}\left<\vert r(t+\Delta t)-r(r)-f(r)\Delta t\vert^2\right>\vert_{r(t)=r}, 
\end{eqnarray}
where $\left<\cdot\right>\vert_{r(t)=r}$ denotes the average over all the increments in trajectories with $r-\Delta r \le r(t)<r+\Delta r$. The reconstructed $f(r)$ and $D(r)$ may depend on the spatial resolution $\Delta t$ and the spatial resolution $\Delta r$. To address this issue, we have tested the dependence with $\Delta r=1, 2, 4$ and $\Delta t=0.1548, 0.3096, 0.4644$ hr (i.e. $\Delta t=1, 2, 3$ frames). As shown in Fig.~\ref{fig:S6-1}, no significant deviation is observed. The reconstruction is robust versus $\Delta r$ and $\Delta t$. The results shown in the main text is with $\Delta r=2$ and $\Delta t=0.1548$ hr. 
\begin{figure}[htbp]
  \centering
  \begin{overpic}[width=0.48\textwidth]{"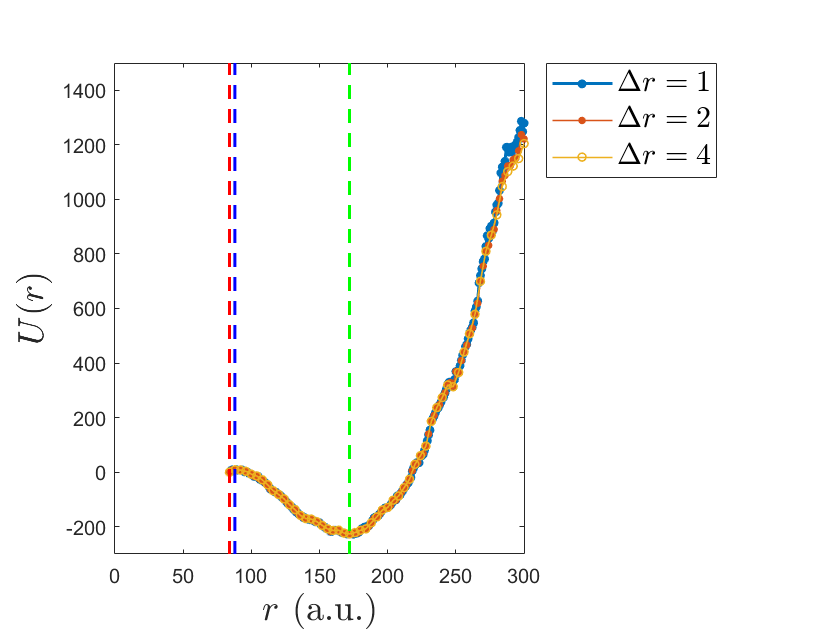"}
    \put(-4,68){\large (a)}
  \end{overpic}
  \begin{overpic}[width=0.48\textwidth]{"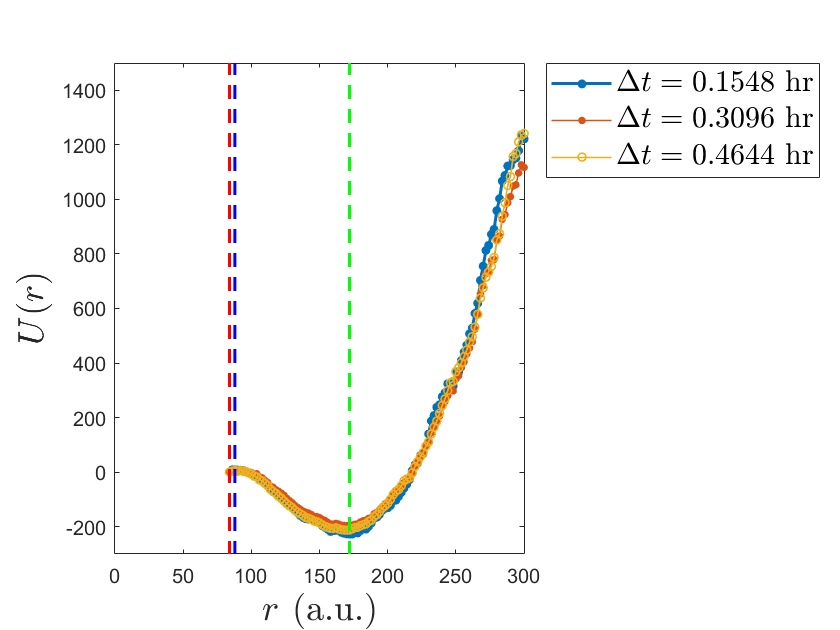"}
    \put(-4,68){\large (b)}   
  \end{overpic}
  \begin{overpic}[width=0.48\textwidth]{"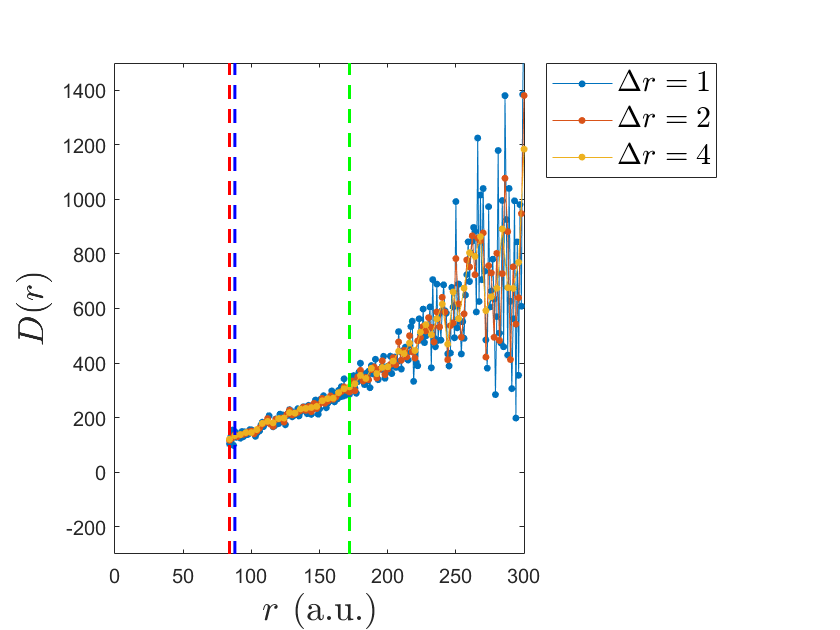"}
    \put(-4,68){\large (c)}  
  \end{overpic}
  \begin{overpic}[width=0.48\textwidth]{"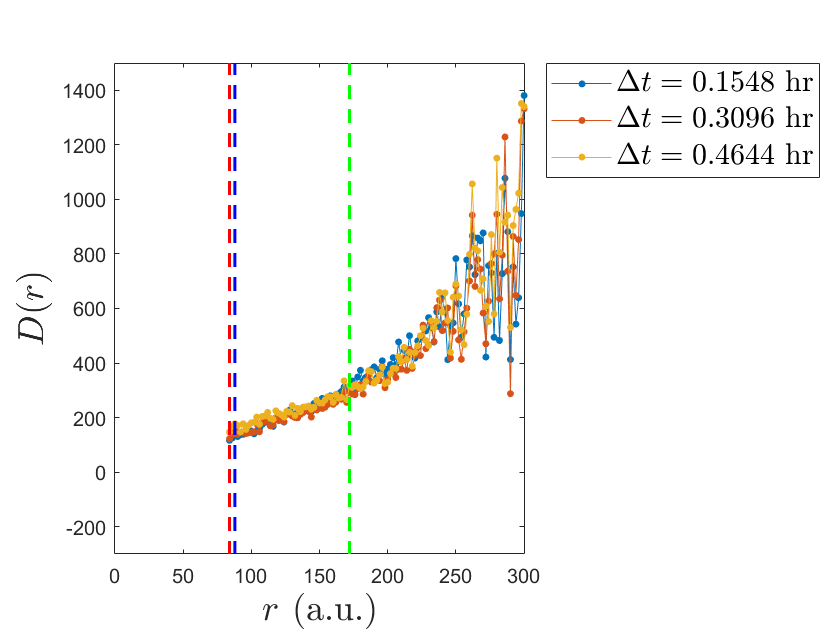"}
    \put(-4,68){\large (d)}  
  \end{overpic}

  \caption{ (a-b) The reconstructed landscape $U(r)=-\int_{r_a}^{r}d\xi f(\xi)$ for various $\Delta r$ (a) and $\Delta t$ (b). (c-d) The reconstructed noise strength $D(r)$ for various $\Delta r$ (c) and $\Delta t$ (d). The colored dashed lines indicate the absorbing boundary  (red), the barrier peak (blue), and the trap center (green). 
  }
  \label{fig:S6-1}
\end{figure}

In this study, we assume the landscape $U(r) = -\int_{r_a}^{r} d\xi f(\xi)$ and noise strength $D(r)$ remain constant throughout the experiment. To test this assumption, we constructed $U$ and $D$ for different time periods. We found that $U$ during the first three hours differed significantly from later periods, likely due to transient effects from initial cell habituation in the mother machine and nutrient switching. As shown in Fig.~\ref{fig:S6-2}, when excluding data from this unstable initial period, the landscape $U$ remains stable, while the noise strength $D$ remains stable throughout the entire experiment.  All analyses presented in the main text and other SI sections are based on data excluding this initial unstable period.
\begin{figure}[htbp]
  \centering
  \begin{overpic}[width=0.48\textwidth]{"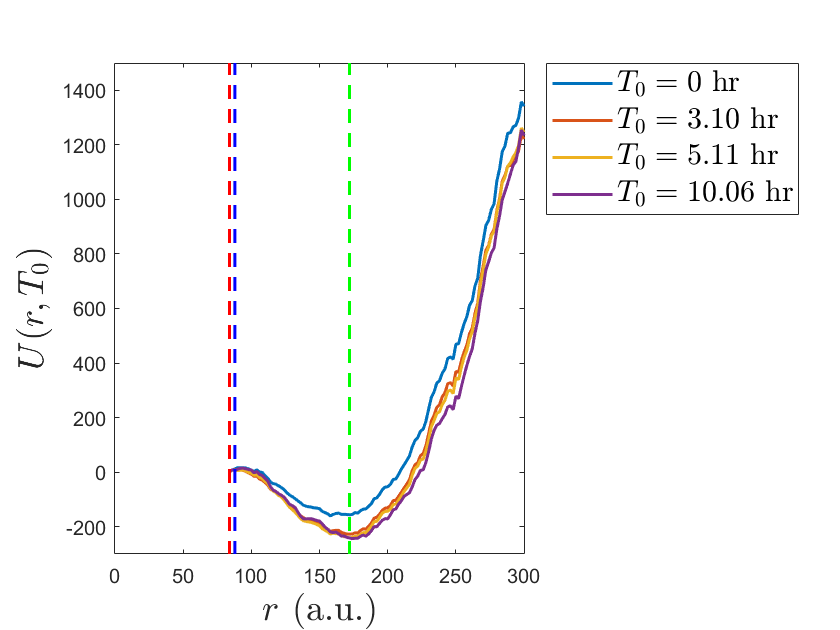"}
    \put(-4,68){\large (a)}
  \end{overpic}
  \begin{overpic}[width=0.48\textwidth]{"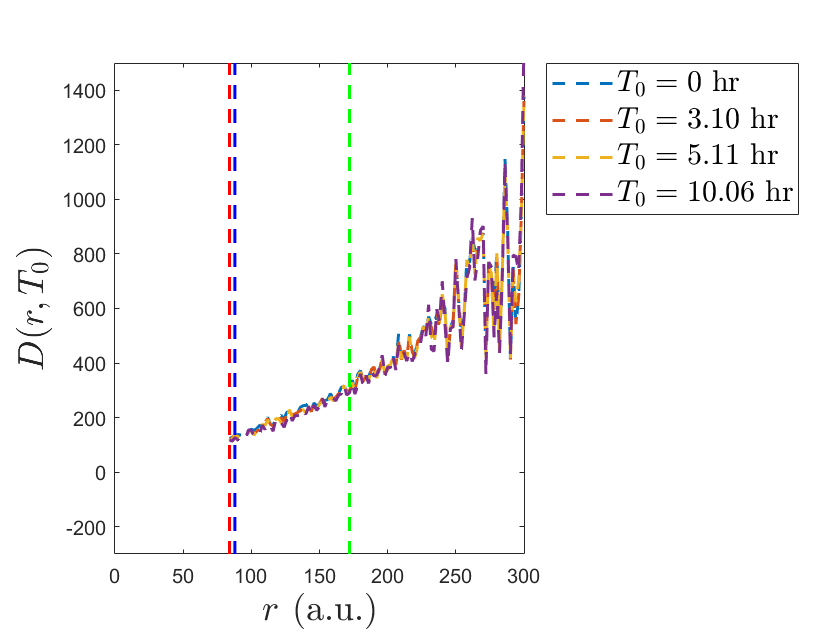"}
    \put(-4,68){\large (b)}   
  \end{overpic}
  \caption{(a) The reconstructed landscape $U(r,T_0)$ truncating the unstable initial period from $T_0=0, 3.096, 5.1084, 10.062$ hr.   (b) Same as (a), but for the noise strength $D(r,T_0)$.  The colored dashed lines indicate the absorbing boundary (red), the barrier peak (blue), and the trap center (green).   }
  \label{fig:S6-2}
\end{figure}

\section{Simulation of the first passage process}
\label{sim}

The state transition is simulated as a first-passage process to the absorbing boundary on the reconstructed landscape $U(r)$ with noise strength $D(r)$. The simulation implements Langevin dynamics equivalent to the Fokker-Planck equation (Eq.~(\ref{fpe}) in the main text and Eq.~(\ref{eq:fpe})). Algorithm~\ref{alg:traj} (pseudocode) summarizes the Euler method used for trajectory generation. 

\begin{minipage}{0.7\textwidth} 
  \begin{algorithm}[H]
    \caption{Langevin simulation of $r(t)$ via Euler method}
    \label{alg:traj}
    \begin{algorithmic}[1]
      \State \textbf{Given:}
      \State \quad Total time $T$ and time step $\Delta t$
      \State \quad Initial condition $r_0 = \text{IC}$
      \State \quad Absorbing boundary $r_a = 84$
      \State \quad Driving force $f(r;\beta)$
      \State \quad Noise strength $D(r;\gamma)$
      \State \quad Normal random generator $N(0,1)$
      \vspace{1mm}
      \State \textbf{Initialize:} $t \gets 0$, $r \gets \text{IC}$
      \vspace{1mm}
      \While{$t < T$}
        \If{$r \ge r_a$}
          \State Generate an increment driven by noise: 
          \[
            \eta \gets \sqrt{2\,D(r;\gamma)} \cdot N\bigl(0,\sqrt{\Delta t}\bigr)
          \]
          \State Update position:
          \[
            r \gets r + \Delta t \cdot f(r;\beta) + \eta
          \]
          \State Advance time: $t \gets t + \Delta t$
        \Else
          \State Mark $r$ as absorbed (e.g., $r \gets -1$)
          \State \textbf{break} \Comment{Exit the loop}
        \EndIf
      \EndWhile
      \vspace{1mm}
      \State \textbf{Output:} The trajectory $r(t)$
    \end{algorithmic}
  \end{algorithm}
\end{minipage}
\vspace{1.5em}

For the simulation, we use the analytical expressions of $U(r)$ and $D(r)$, which are fitted to their discrete versions constructed from experimental data. We chose a Fourier series as the fitting function for $U$ in the range $r \in [84,250]$, 
\begin{equation}
U_{\text{fit}}(r) = a_0 + \sum_{n=1}^{3} \left[ a_n \cos(n \omega r) + b_n \sin(n \omega r) \right]
\label{eq:7-1}
\end{equation}
The coefficients $ a_n$, $b_n$, and $\omega$ were determined using MATLAB's Curve Fitting Toolbox (cftool) with a trust-region-based nonlinear least-squares algorithm. The drift force was then estimated as $f_{\text{fit}}(r)=-\partial U_{\text{fit}}/\partial r$. For $r > 250$, where experimental data provided limited information, we extended $f_{\text{fit}}(r)$ linearly as $ f_{\text{fit}}(r) = kr + b $. The parameters $k$ and $b$ were chosen to ensure smoothness at $r = 250$.

We also simulated the process on modified landscapes with scaled heights to investigate the influence of the barrier height. The modified drift force is defined as
\begin{equation}
f(r;\beta) = f_{\text{fit}}(r)\left[\beta H(r_c - r) + H(r - r_c)\right],
\label{eq:7-2}
\end{equation}
where $H(\cdot)$ is the Heaviside step function, $r_c$ denotes the trap center, and $\beta$ modulates the barrier height. The modified landscape $U$ can be obtained as
\begin{equation}
U(r; \beta) = -\int f(r; \beta), dr + C,
\label{eq:7-3}
\end{equation}
where the integration constant $C$ is chosen such that all curves with different $\beta$ coincide at $r = r_c$. Thus, $\beta = 1$ reproduces the original fitted landscape, $\beta = 0$ flattens the landscape, and larger values (e.g., $\beta = 7$) elevate the barrier (see Fig.~\ref{fig3}(a) in the main text).

The noise strength $D$ is fitted to a cubic polynomial:
\begin{equation}
D_{\text{fit}}(r) = \sum_{n=0}^{3} c_n r^n,
\label{eq:7-4}
\end{equation}
using MATLAB’s Curve Fitting Toolbox.
To investigate the influence of noise on the transition process, we introduce a modified noise strength based on Eq.~(\ref{dgamma}) in the main text:
\begin{equation}
D(r; \gamma) = D_{\text{fit}}(r_c) + \gamma \left[ D_{\text{fit}}(r) - D_{\text{fit}}(r_c) \right].
\label{eq:7-5}
\end{equation}
Here, $\gamma = 1$ corresponds to the experimentally fitted, position-dependent noise strength, while $\gamma = 0$ yields a spatially uniform noise profile $D(r) = D_{\text{fit}}(r_c)$ (see Fig.~\ref{fig3}(b) in the main text).

The fitting parameters for both $ U_{\text{fit}}(r) $ and  $ D_{\text{fit}}(r) $ are given in Table~\ref{tab:table7}.

\begin{table*}[htbp]
\caption{\label{tab:table7} Fitting parameters for the landscape $U_{\mathrm{fit}}(r)$ and noise strength $D_{\mathrm{fit}}(r)$.}
\begin{ruledtabular}
\begin{tabular}{cccc}
\multicolumn{4}{c}{Landscape $U_{\mathrm{fit}}(r)$; Linear Extension of $f_{\mathrm{fit}}(r)$ for $r>250$} \\
\colrule
Parameter & Value & Parameter & Value \\
$a_{0}$ & $-2.71597638837018\times 10^{12}$ & $b_{1}$ & $-2.43595049642955\times 10^{11}$ \\
$a_{1}$ & $4.06666119464276\times 10^{12}$ & $b_{2}$ & $1.94520725495797\times 10^{11}$ \\
$a_{2}$ & $-1.61791120033713\times 10^{12}$ & $b_{3}$ & $-4.84823435528684\times 10^{10}$ \\
$a_{3}$ & $2.67226388294186\times 10^{11}$ & $\omega$ & $-3.31103303639035\times 10^{-4}$ \\
$k$ & $-1.62054558396339\times 10^{1}$ & $b$ & $-1.56089663505554\times 10^{-1}$ \\
\colrule
\multicolumn{4}{c}{Noise $D_{\mathrm{fit}}(r)$} \\
\colrule
Parameter & Value & Parameter & Value \\
$c_{0}$ & $1.61622336564688\times 10^{1}$ & $c_{2}$ & $-2.30433321514461\times 10^{-4}$ \\
$c_{1}$ & $1.19131489155714$ & $c_{3}$ & $1.99731573774679\times 10^{-5}$ \\
\end{tabular}
\end{ruledtabular}
\end{table*}

To evaluate the MFPT depending on the initial state, the Langevin dynamics is simulated starting from the initial site $r_0\in[84,250]$ with increments of $dr_0=2$. For each initial site, $N=10^5$ particles are simulated. The total simulation time is large enough to ensure all the particles reached the absorbing boundary at $r_a=84$. It leads to Fig.~\ref{fig4} in the main text. 

Here we further show the distributions of first passage times (FPT) for several representative initial conditions $r_0$ under $\beta=1$ and $\gamma=1$. As shown in Fig.~\ref{fig:S7-2}, the FPT distributions depend strongly on the initial condition. The semilog curves in Fig.~\ref{fig:S7-2}(b) resemble the compact-case distributions shown in Fig.~3 of Ref.~\cite{benichou2010geometry}, consistent with the picture described by Eq.~(4) of Ref.~\cite{benichou2010geometry}, where the FPT distribution contains two components: trajectories moving directly toward the absorbing boundary, and trajectories that first undergo effective reflection from more distant positions before eventual absorption. By contrast, in the Kramers regime the FPT distribution is dominated by a single exponential. The distributions shown here therefore indicate non-Kramers transitions.
\begin{figure}[htbp]
  \centering
  \begin{overpic}[width=0.49\textwidth]{"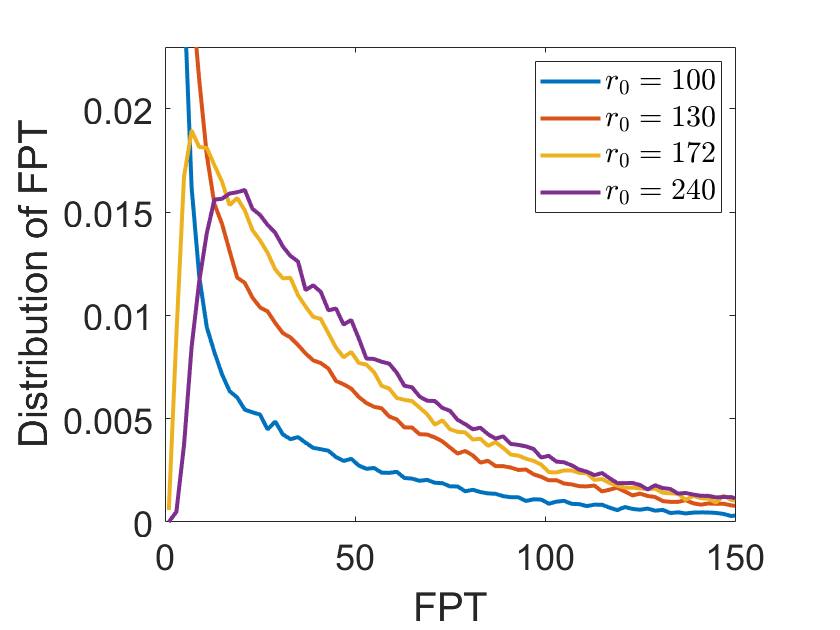"}
    \put(-4,68){\large (a)}
  \end{overpic}
  \begin{overpic}[width=0.49\textwidth]{"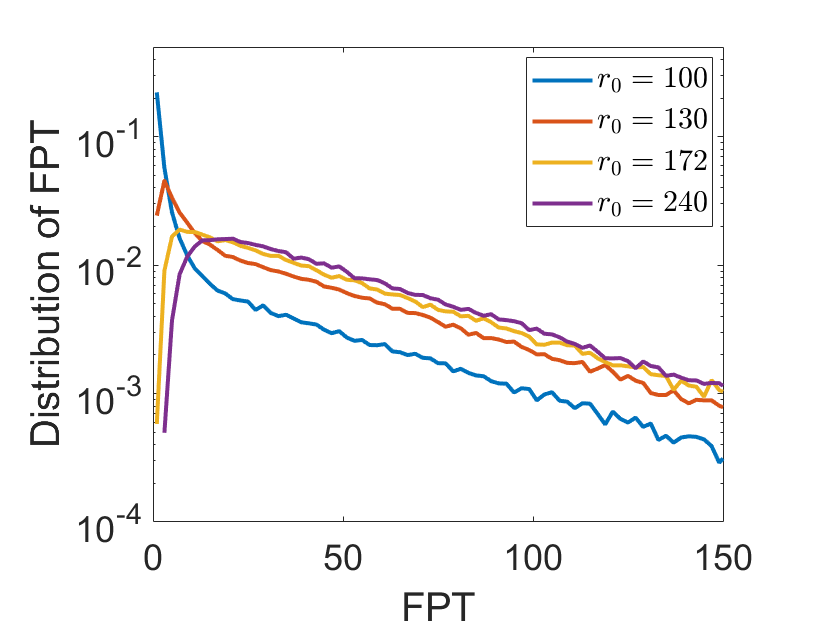"}
    \put(-4,68){\large (b)}   
  \end{overpic}
  \caption{First-passage time distributions $P(\tau \mid r_0)$ for several representative initial states $r_0$ under $\beta=1$ and $\gamma=1$, shown on (a) linear and (b) semi-logarithmic scales.}
  \label{fig:S7-2}
\end{figure}

\FloatBarrier 

\afterpage{%
  \clearpage
  \pagenumbering{arabic}%
}
\clearpage
\section*{Additional Supplementary Figures}

\begin{figure}[htbp]
  \centering
  \begin{overpic}[width=0.99\textwidth]{"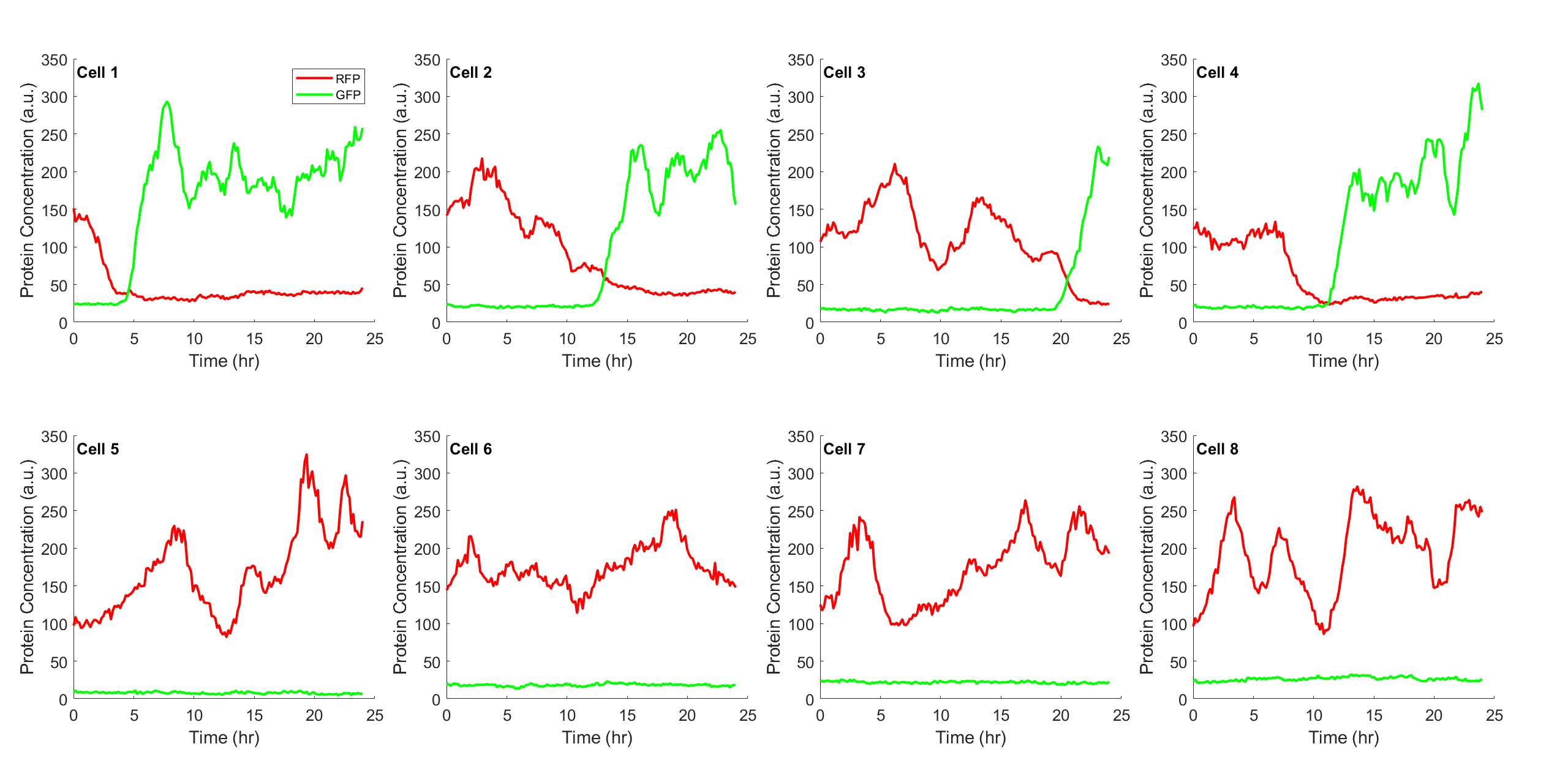"}
    \put(-2.6,47){\large(a)}
  \end{overpic}\\[1em]
  \begin{overpic}[width=1\textwidth]{"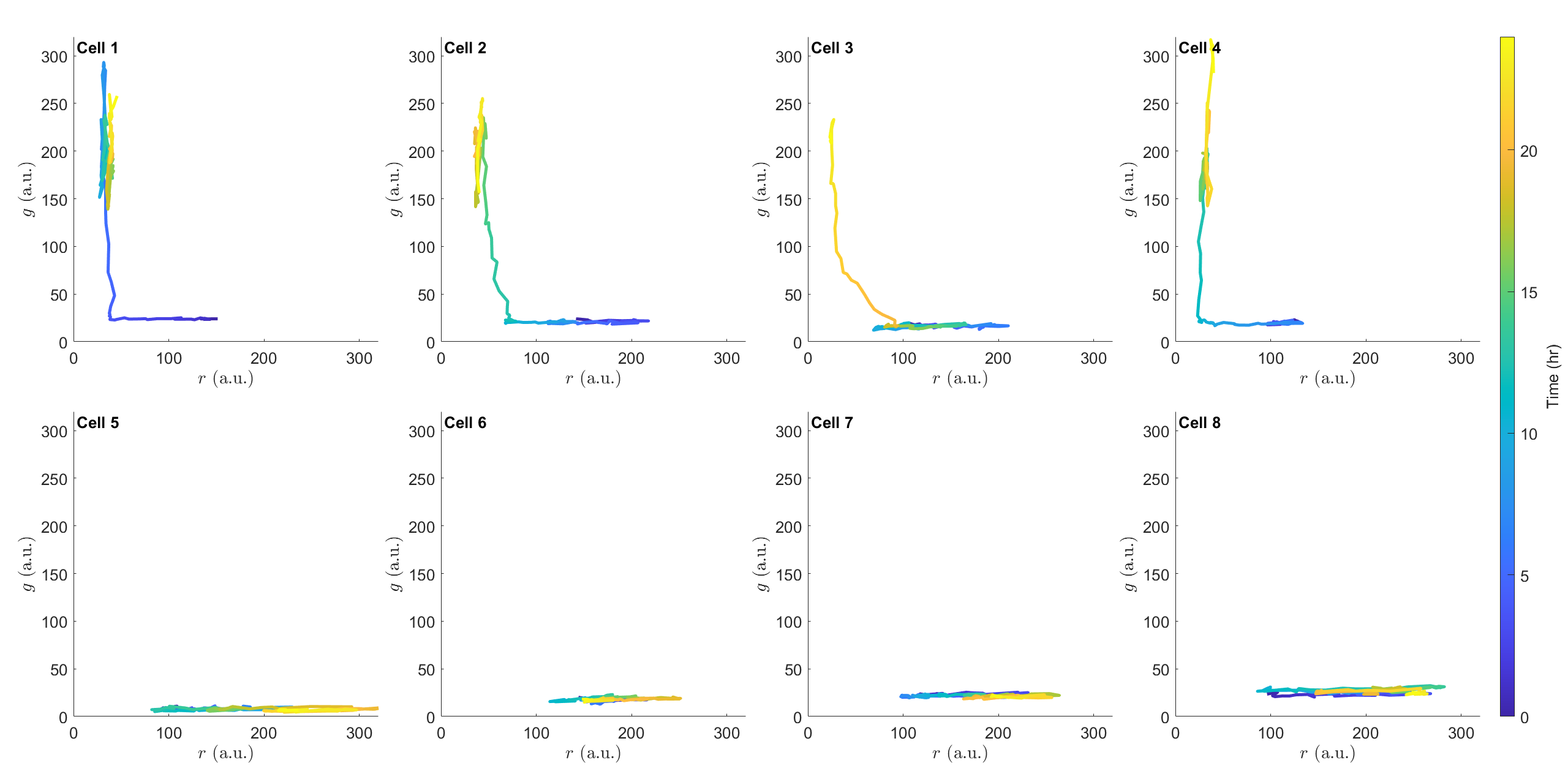"}
    \put(-2,47){\large(b)}   
  \end{overpic}
  \caption{Typical trajectories of single cell fluorescent intensity. 
  (a, Row 1-2) Time series of RFP (red curves) and GFP (green curves) intensities. Row 1 shows typical transition processes. Row 2 shows those for cells remain in the R-state throughout the whole experiment. 
  (b, Row 3-4) The above trajectories plotted in the $(r, g)$ plane. The color indicates time-lapse. 
  }
  \label{fig:S4-1}
\end{figure}

\begin{figure}[htbp]
  \centering
  \begin{overpic}[width=0.48\textwidth]{"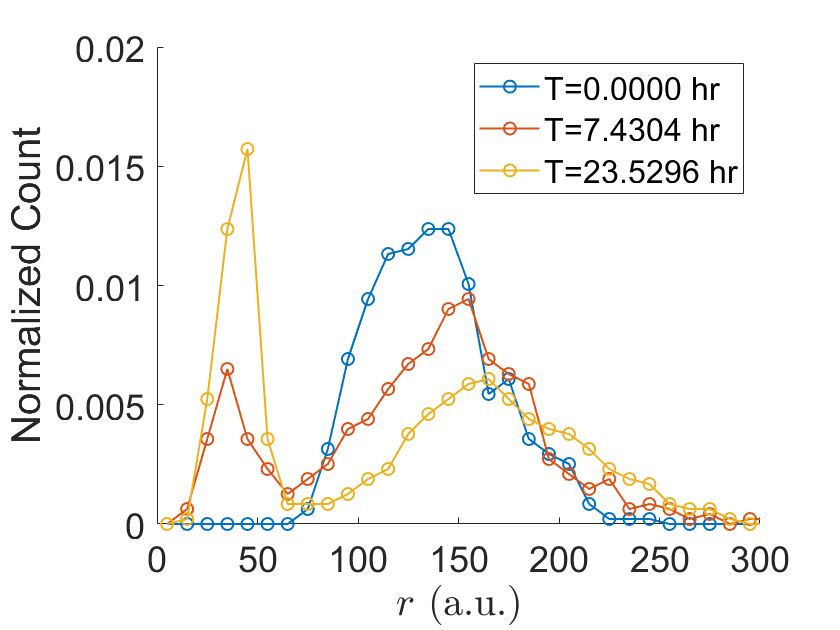"}
    \put(-4,68){\large (a)}
  \end{overpic}
  \begin{overpic}[width=0.48\textwidth]{"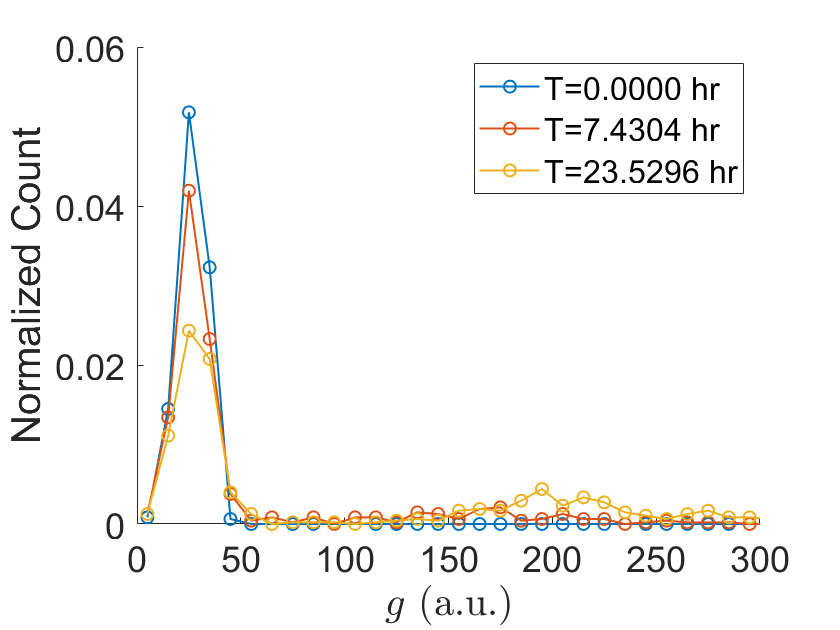"}
    \put(-4,68){\large (b)}   
  \end{overpic}
  \caption{Temporal evolution of fluorescence intensity distributions. 
Probability density function of RFP (a) and GFP (b) intensities at three representative time: \( t = 0 \,\text{hr} \) (blue dot line), \( 7.43 \,\text{hr} \) (red dot line), and \( 23.53 \,\text{hr} \) (yellow dot line). 
  }
  \label{fig:S4-2}
\end{figure}

\begin{figure}[htbp]
  \centering
  \includegraphics[width=1.05\textwidth]{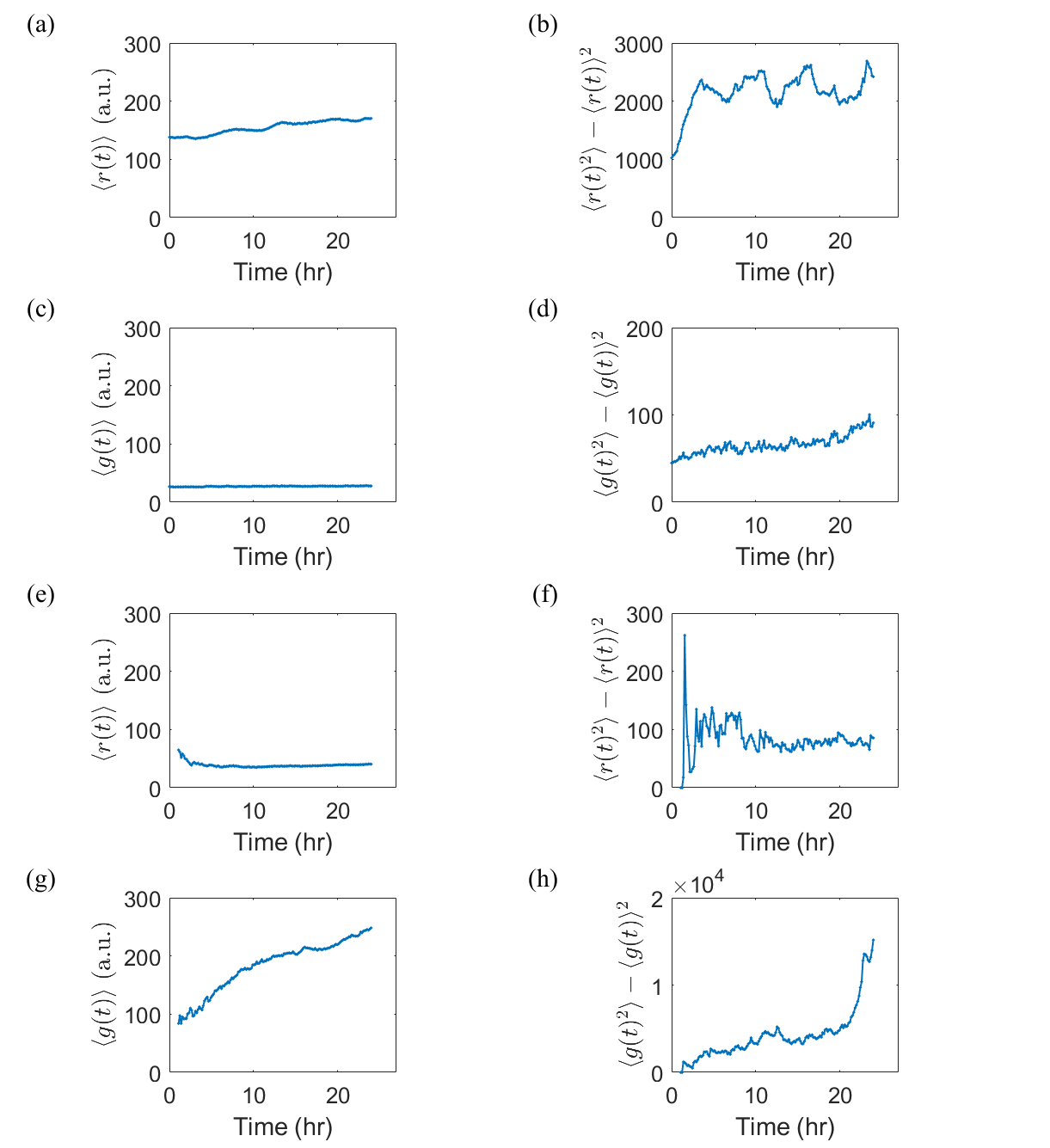}
  \caption{
  Temporal evolution of fluorescence intensity for R-state and G-state cells. 
  Panels (a–d) show the mean (left) and variance (right) of RFP (Row 1) and GFP (Row 2) intensities for cells that remain in the R-state throughout the whole experiment. 
  Panels (e–h) are the same with Panel (a-d), but for cells that transition to the G-state. The continuous evolution in G-state is observed (see Row 4). 
  For cells remaining in the R-state, the variance of $g$ stays small (Panel (d)), suggesting that the contribution from averaging over the $g$-direction could be negligible. This supports approximating the R-state first-passage dynamics by an effective one-dimensional description along $r$.}
  \label{fig:S4-3}
\end{figure}
\end{onecolumngrid}

\end{document}